\documentclass{gji}
\usepackage{timet,color}
\usepackage[urlcolor=blue,citecolor=black,linkcolor=black]{hyperref}
\usepackage{graphicx}
\usepackage{epstopdf}
\usepackage{amsmath}
\usepackage{xcolor}

\usepackage{xparse,xcoffins}

\ExplSyntaxOn
\NewCoffin\imagecoffin
\NewCoffin\labelcoffin

\keys_define:nn { miguel/label }
 {
  label   .tl_set:N = \l_miguel_label_tl,
  labelbox .bool_set:N = \l_miguel_label_box_bool,
  labelbox .default:n = true,
  fontsize .tl_set:N = \l_miguel_label_size_tl,
  fontsize .initial:n = \footnotesize,
  pos .choice:,
  pos/nw .code:n = \tl_set:Nn \l_miguel_label_pos_tl { left,up },
  pos/ne .code:n = \tl_set:Nn \l_miguel_label_pos_tl { right,up },
  pos/sw .code:n = \tl_set:Nn \l_miguel_label_pos_tl { left,down },
  pos/se .code:n = \tl_set:Nn \l_miguel_label_pos_tl { right,down },
  pos/n .code:n = \tl_set:Nn \l_miguel_label_pos_tl { hc,up },
  pos/w .code:n = \tl_set:Nn \l_miguel_label_pos_tl { left,vc },
  pos/s .code:n = \tl_set:Nn \l_miguel_label_pos_tl { hc,down },
  pos/e .code:n = \tl_set:Nn \l_miguel_label_pos_tl { right,vc },
  pos .initial:n = nw,
  unknown .code:n   = \clist_put_right:Nx \l_miguel_label_clist
                       { \l_keys_key_tl = \exp_not:n { #1 } }
 }
\clist_new:N \l_miguel_label_clist
\box_new:N \l_miguel_label_box
\box_new:N \l_miguel_label_image_box

\NewDocumentCommand{\xincludegraphics}{O{}m}
 {
  \group_begin:
  \tl_clear:N \l_miguel_label_tl
  \clist_clear:N \l_miguel_label_clist
  \keys_set:nn { miguel/label } { #1 }
  \tl_if_empty:NTF \l_miguel_label_tl
   {
    \miguel_includegraphics:Vn \l_miguel_label_clist { #2 }
   }
   {
    \SetHorizontalCoffin\imagecoffin
     {
      \miguel_includegraphics:Vn \l_miguel_label_clist { #2 }
     }
    \SetHorizontalCoffin\labelcoffin
     {
      \raisebox{\depth}
       {
        \bool_if:NTF \l_miguel_label_box_bool
         { \fcolorbox{white}{white}{\l_miguel_label_size_tl\l_miguel_label_tl} }
         { \l_miguel_label_size_tl\l_miguel_label_tl }
       }
     }
    \SetVerticalPole\imagecoffin{left}{3pt+\CoffinWidth\labelcoffin/2}
    \SetVerticalPole\imagecoffin{right}{\Width-3pt-\CoffinWidth\labelcoffin/2}
    \SetHorizontalPole\imagecoffin{up}{\Height-3pt-\CoffinHeight\labelcoffin/2}
    \SetHorizontalPole\imagecoffin{down}{3pt+\CoffinHeight\labelcoffin/2}
    \use:x{\JoinCoffins\imagecoffin[\l_miguel_label_pos_tl]\labelcoffin[vc,hc]} 
    \TypesetCoffin\imagecoffin
   }
   \group_end:
 }
\NewDocumentCommand{\setlabel}{m}
 {
  \keys_set:nn { miguel/label } { #1 }
 }

\cs_new_protected:Nn \miguel_includegraphics:nn
 {
  \includegraphics[#1]{#2}
 }
\cs_generate_variant:Nn \miguel_includegraphics:nn { V }

\ExplSyntaxOff

\title[MUYSC: An end-to-end muography simulation toolbox]
  {MUYSC: An end-to-end muography simulation toolbox}
  
\author[J. Peña-Rodríguez et. al]
  {J. Peña-Rodríguez$^{1,2}$, J. Jaimes-Teherán$^3$, K. Dlaikan-Castillo$^3$,  and   L.A. Núñez$^{2,4}$ \\
  $^1$ Fakultät für Mathematik und Naturwissenschaften, Bergische Wuppertal Universität, Wuppertal, Germany \\
  $^2$ Escuela de Física, Universidad Industrial de Santander, Bucaramanga, Colombia \\
  $^3$ Escuela de Ingeniería de Sistemas e Informática, Universidad Industrial de Santander, Bucaramanga, Colombia \\
  $^4$ Departamento de Física, Universidad de Los Andes, Mérida Venezuela
  }

\date{today}
\pagerange{\pageref{firstpage}--\pageref{lastpage}}
\volume{200}
\pubyear{1998}

\begin{document}

\label{firstpage}

\maketitle

\begin{summary}
Muography is an imaging technique based on attenuation of the directional muon flux traversing geological or anthropic structures. Several simulation frameworks help to perform muography studies by combining specialised codes from the muon generation (CORSIKA and CRY) and the muon transport (GEANT4, PUMAS, and MUSIC) to the detector performance (GEANT4). This methodology is very precise but consumes significant computational resources and time.

In this work, we present the end-to-end python-based \textbf{MU}ograph\textbf{Y} \textbf{S}imulation \textbf{C}ode. MUYSC implements a muography simulation framework capable of rapidly estimating rough muograms of any geological structure worldwide. MUYSC generates the muon flux at the observation place, transports the muons along the geological target, and determines the integrated muon flux detected by the telescope. Additionally, MUYSC computes the muon detector parameters (acceptance, solid angle, and angular resolution) and reconstructs the 3-dimensional density distribution of the target. We evaluated its performance by comparing it with previous results of several simulation frameworks. 
\end{summary}

\begin{keywords}
 Muography Simulation Toolbox; Muon Tomography; Muon Radiography; Muon Telescope
\end{keywords}

\section{Introduction}

\label{int}

Muography is a non-invasive technique used to explore geological or anthropic structures acquiring two-dimensional density distributions by registering the integrated muon flux passing through the target. Muon tomography can reconstruct three-dimensional density distributions combining multi-projection muography data, considering the observation site, the screening of neighbourhood mountains and the muon detector parameterization.

Muon tomography reconstructs three-dimensional density distributions of the target from two-dimensional density projections. There exist two main techniques: Inversion methods and filtered-back projections. Inversion methods combine muography with gravimetry data \cite{Tanaka2010,Nishiyama2014,Guardincerri2017,Vesga2021}. Barnound and collaborators inverted gravimetric and muographic data of the Puy de Dôme volcano. They conclude that muographic data cannot retrieve a 3D distribution of the volcano from a single observation point - It's necessary to combine it with gravimetric data \cite{Barnoud2021}. Marteau and collaborators combined muography information (3 and 5 telescopes) with gravity data of the La Soufriére de Guadeloupe lava dome. The results showed density anomalies inside the lava dome with good agreement with electrical tomography data \cite{Marteau2017,RosasCarbajal2017}.

Filtered Back Projection methods use several muon radiography recordings around the target. Nagahara and collaborators performed a muon tomography of the Omuroyama volcano with an internal density distribution similar to a checkerboard. The reconstruction errors depend on the number of observation points (they evaluated 4, 8, 16, 32, and 64 points) and the exposure time~\cite{Nagahara2018}.

Different software tools form composite simulation frameworks in muography: from cosmic ray showers to detector response. In the Monte Carlo approach, authors use specialised codes such as CORSIKA (COsmic Ray SImulations for KAscade), CRY, ARTI, and EcoMug~\cite{Moussawi2022,Sarmiento2022,Samalan2022,Taboada2022}. CORSIKA simulates the evolution of extensive air showers in the atmosphere step-by-step, giving high accuracy in particle spectrum composition. Its primary disadvantage is the high simulation time (hours or even days to compute the differential muon flux at the observation site). CRY contains parametrizations of all particles, including background, with execution times of the order of minutes. EcoMug generates cosmic muons according to a differential flux characterisation provided by a user or with a default model based on data from the ADAMO experiment. Its execution times are in the order of seconds. Semi-empirical models offer a fast methodology to compute muon fluxes. They fit muon flux models considering the zenithal incoming angle, muon energy, and altitude, obtaining comparable results with Monte Carlo-based simulation frameworks \cite{Lechmann2021,Lesparre2010}.

Other Monte Carlo methods (GEANT4, PUMAS, and MUSIC) simulate the muon-target interaction: GEANT4 provides accurate muon tracking and geometry/material target structure, allowing the detailed study of the muon scattering effects. PUMAS library is used for transporting muon and tau leptons in matter. It operates in a fast-deterministic or detailed Monte Carlo mode~\cite{Niess2022}. MUSIC integrates energy the losses of muons due to: ionization, bremsstrahlung, electron-positron pair production and muon-nucleus inelastic scattering \cite{Kudryavtsev2009}. Semi-empirical methods perform muon transport employing models that parameterize the stopping power of muons in different materials based on the Groom dataset~\cite{Groom2001,Lesparre2010}.

\begin{figure*}
\begin{center}
\vbox to100mm{\vfil
\includegraphics[width=0.8\textwidth]{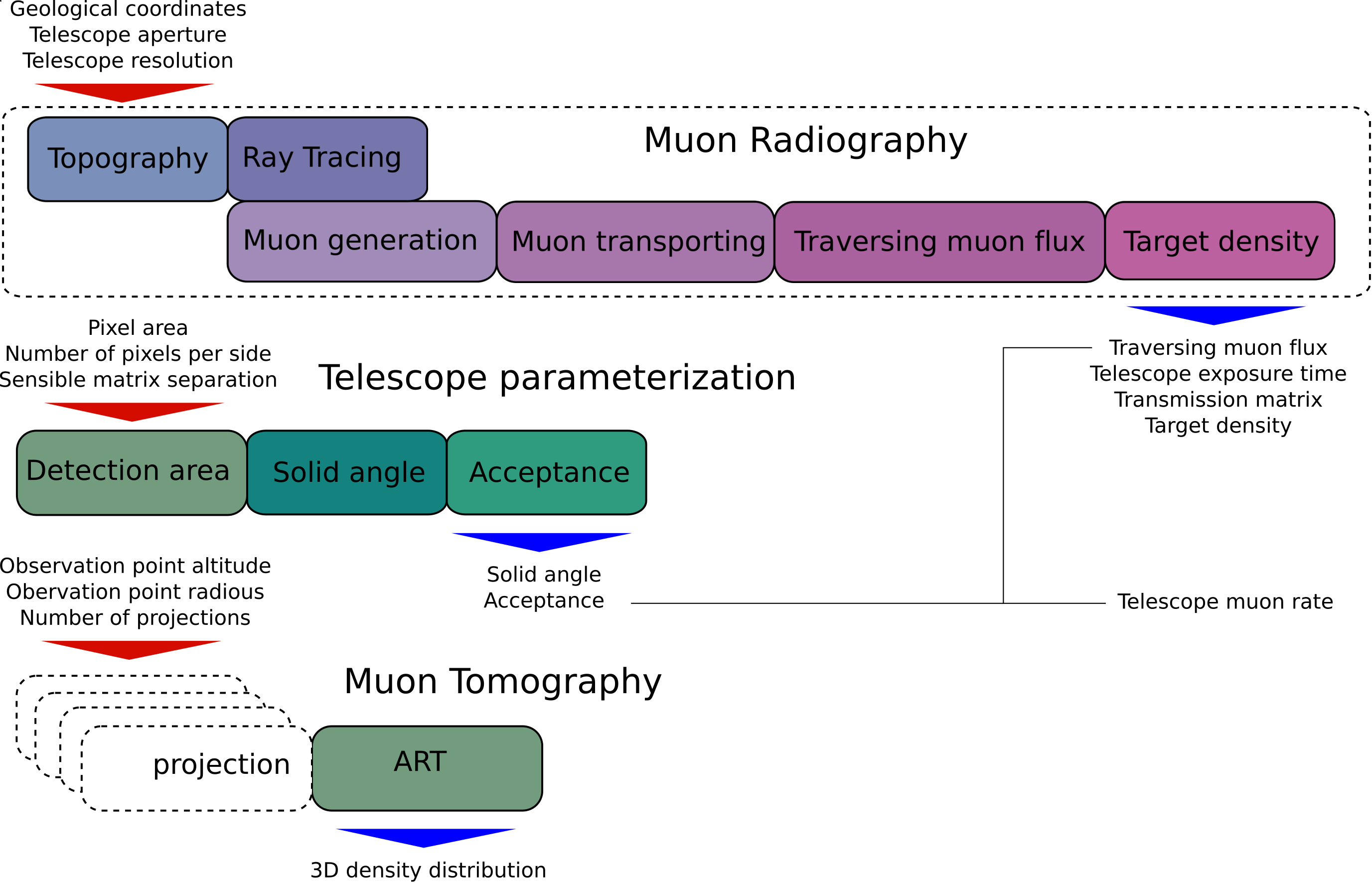}
\vfil}
\caption{Workflow of the MUYSC framework with three modules: muon radiography, telescope parameterization, and muon tomography. The solid line boxes represent the sub-modules within each module. The red arrows represent intakes, the blue ones indicate the output and the dashed line tomography box estimates each projection. The muon tomography module reconstructs a three-dimensional density distribution of the target from several (simulated or measured) muon radiography projections. This module loads the geological coordinates of the target topography, the observation point coordinates, the projection point coordinates, the telescope angular aperture (zenith and azimuth), and the resolution. The telescope parameterization module calculates the detector acceptance, solid angle, and detection area from the pixel area, the number of pixels and the separation distance between detection matrices.}
\label{fig::muysc}
\end{center}
\end{figure*}

The muon detector's performance depends on geometrical and functional features: acceptance, solid angle, and angular resolution, characterising the overall muon detector performance. These geometrical parameters depend on the pixel area, the number of pixels and the distance between the detection layers \cite{Lesparre2010}. Generally, muography experiments address the detector response through GEANT4 simulations, providing information about the energy and time-of-flight of muons crossing the detector. GEANT4 allows us to create sensitive detector matrices with specific materials and photo-detector quantum efficiencies. We also can evaluate the installation of passive layers (lead or iron) for background filtering \cite{VsquezRamrez2020,Moussawi2022,Samalan2022}. Semi-empirical models deal with detector response simulations based on data collected from an operating detector. These models include sensor features (photo-electron spectrum, gain, temperature dependency) and signal features (attenuation, delay, and noise) \cite{PenarodriguezEng2021}.

This paper introduces MUYSC (MUographY Simulation Code), a python-based framework for simulating muon radiography (atmospheric muon generation, muon transport, and geometrical detector parameterization) and muon tomography (three-dimensional reconstruction) of geological targets.  In section \ref{chap::radiography}, we describe the muon radiography method used by MUYSC, while section \ref{chap::tomography} displays the methodology employed by the MUYSC muon tomography module. Section \ref{chap::telescope} is about how MUYSC computes the muon detector parameters. Finally, in section \ref{chap::conclusions}, we wrap up some concluding remarks. 

\section{MUYSC framework}
\label{chap::musyc}

MUYSC is an open-source, python-based code with three main functions as shown in Figure \ref{fig::muysc}:
\begin{itemize}
    \item Muon radiography
    \item Telescope parameterization
    \item Muon tomography
\end{itemize}
The muon radiography module inputs are the geological coordinates of the target topography, the observation point coordinates, the projection point coordinates, the telescope angular aperture (zenith and azimuth), and the resolution. It uploads the topography data of the geological target, sets the observation point and identifies the projection planes where the telescope is pointing. By ray tracing, it estimates the traversing distances of muons inside the object from all the possible trajectories to the telescope, considering the angular aperture and resolution. Semi-empirical models estimate the differential muon flux depending on the zenith angle and the observation point altitude. Then, an algorithm estimates the minimum energy required by the muon for traversing a given distance inside the target. With this minimum energy and the muon incidence angle, we calculate the integrated muon flux, the telescope exposure time (assuming an expected muon rate per pixel), the target transmission matrix, and the density.

The telescope parameterization module calculates the detector acceptance, solid angle, detection area (from the pixel area), the number of pixels and the separation distance between the detection matrices. This module also estimates the expected muon rate using the integrated particle flux crossing the target and the telescope acceptance.

The muon tomography module reconstructs a three-dimensional density distribution of the target from several (simulated or measured) muon radiography projections, having the observation points as inputs.

\section{Muon radiography}
\label{chap::radiography}
This module performs muon radiography of any geological target by inputting the topography, observation point and projection point, angular aperture and resolution of the detector. In this section, we describe how MUYSC creates the target topography, the muon flux generation by using semi-empirical models, and the muon propagation algorithm.

\subsection{Topography data}

The topography script loads the volcano topography and converts the global coordinates (latitude, longitude, altitude) into a local metric system. The module also locates the observation and projection points.

MUYSC downloads the target topography from the SRTM ({\it Shuttle Radar Topography Mission}) NASA dataset at a resolution of 1 arc-second (30 meters) \cite{strm}. The dataset contains the latitude, longitude, and altitude per point. The script inputs the target area coordinates ($Long_{max}$, $Long_{min}$, $Lat_{max}$, $Lat_{min}$). Then, the user establishes the observation point (coordinates of the muon telescope) and the projection point (coordinates where the telescope is pointing).

MUYSC converts global coordinates (Latitude, longitude) to cartesian coordinates ($x,y$ in meters) as follows,
\begin{equation}
x = C_e*(Long_0-Long_i)* \frac{\pi * \cos (Lat_0+Lat_i)*360^{\circ}}{180^{\circ}},
\end{equation}
\begin{equation}
y = C_e*\frac{(Lat_0-Lat_i)}{360^{\circ}},
\end{equation}
where ($Long_0$, $Lat_0$) are the (longitude, latitude) of the reference point, and $C_e$ is the Earth's circumference.

\subsection{Ray tracing}
MUYSC estimates the muon travelling distance along the target for every azimuth and zenith angle depending on the detector aperture and angular resolution. The algorithm traces a line joining the observation point ($x_{op}, y_{op}, z_{op}$) with the projection point observation point ($x_{pp}, y_{pp}, z_{pp}$), and obtains the altitude for $P$ points along the projected line. MUYSC evaluates if the altitude of the $i$th point is above the projected line. If it is, the distance between the ($i-1$)th and $i$th point is added to the total traversing distance. We get a travelling distance matrix for $(2N-1)^2$ trajectories, where $N$ is the number of scintillator strips.


\subsection{Muon flux models}
MUYSC estimates the incident muon flux on the target using semi-empirical models that reproduce results of Monte Carlo methods such as CRY~\cite{Su2021}.

\subsubsection{Gaisser model}
The Gaisser parameterization model of the differential muon flux in $[cm^{-2}sr^{-1}s^{-1}GeV^{-1}]$ units and is follows~\cite{Gaisser1990,Lesparre2010}:
\begin{equation}
    \Phi_G( E_{0}, \theta) = A_G E_0^{-\gamma} \left( \frac{1}{1+\frac{\hat{E}_{0} \cos \theta}{E_{0,\pi}}} + \frac{B_G}{1+\frac{\hat{E}_{0} \cos \theta}{E_{0,k}}} + r_c \right), 
\end{equation}
where $A_G$ is a scale factor, $\gamma$ the power index, $B_G$ a balance factor, which depends on the ratio of muons produced by pions and kaons, $r_c$ the ratio of the prompt muons produced by the decay of charmed particles, and $\theta$ the muon incidence zenith angle. $E_0$ represents the energy of muons at sea level and $\hat{E}_0$ on top of the atmosphere ($E_0 \approx \hat{E}_0$). $E_{0,\pi}$ and $E_{0,K}$ are interpreted as the critical energies of pions and kaons for vertical incidence (i.e. $\theta=0$). Most authors consider them adjustable parameters as shown in Table \ref{tab::gaisser}. MUYSC uses Gaisser's parameters.

\begin{table*}
\begin{minipage}{106mm}
\caption{Parameters of the Gaisser's model for muon generation determined by several authors.}
\label{tab::gaisser}
\begin{tabular}{cccccccc}
\hline
Parameters  & $A_G$ & $B_G$  & $\gamma$ & $E_{0,\pi}$  & $E_{0,K}$  & $r_c$  & $E_0$ range (GeV) \\ \hline
Volkova     & 0.1258    & 0.0588    & 2.65      & 100       & 650       & 0  & 100 - 10$^5$ \\
Gaisser     & 0.14      & 0.054     & 2.70      & 115/1.1   & 850/1.1   & 0  & 100 - 10$^5$ \\
Klimushin   & 0.175     & 0.037     & 2.72      & 103       & 810       & 0  & 300 - 2.5$\times 10^5$ \\
Aglietta    & 0.256     & 0.054     & 2.77      & 115/1.1   & 850/1.1   & 0  & 2$\times 10^3$ - 4$\times 10^4$ \\
Ambrioso    & 0.26      & 0.054     & 2.78      & 115/1.1   & 850/1.1   & 0  & 500 - 2$\times 10^4$ \\
\hline
\end{tabular}
\end{minipage}
\end{table*}

\subsubsection{Gaisser-Tang model}
In 2006 Tang et al. \cite{Tang2006} introduced a modified version of the original Gaisser model taking into account an overestimation of the incident muon flux at low energies ($E_0 < 100 \cos \theta$~GeV):
\begin{equation}
    \Phi_T( E_{0}, \theta) = A_T E_0^{-\gamma} \left( \frac{1}{1+\frac{\hat{E}_{0} \cos \theta^*}{E_{0,\pi}}} + \frac{B_G}{1+\frac{\hat{E}_{0} \cos \theta^*}{E_{0,k}}} + r_c \right),
\end{equation}
where,
\begin{equation}
    \cos \theta^* = \sqrt{ 1 - \frac{1 -\cos^2 \theta}{\left( 1 + \frac{H_{atm}}{R_{Earth}} \right)^2}},
\end{equation}
where $R_{Earth}$ = 6370\,km is Earth's radius and $H_{atm}$ = 32\,km is the altitude where muons with large angle trajectories are produced~\cite{Lesparre2010}. The muon energy is,
\begin{equation}
    \hat{E}_0 = E_0 + \Delta E_0,
\end{equation}
where,
\begin{equation}
    \Delta E_0 = 0.00206 \left( \frac{1030}{\cos \theta^*} - 120 \right),
\end{equation}
and, 
\begin{equation}
    AT = AG \left( \frac{120\cos \theta^*}{1030} \right)^{\frac{1.04}{(E_0+\Delta E_0/2)\cos \theta^*}}.
\end{equation}
MUSIC uses a version of this modified spectrum with $r_c = 10^{-4}$ \cite{Kudryavtsev2009}.

\subsubsection{Bugaev model}
Several authors proposed an empirical model based on muon flux measurements at sea level \cite{Bugaev1998,Lesparre2010,Lechmann2021,Su2021}. The model is expressed in the form of a fitting formula, 
\begin{equation}
    \Phi_B(p) = A_B p^{-(a_3y^3+a_2y^2+a_1y+a_0)},
\end{equation}
where $y = \log_{10} p$ and the muon momentum $p$ in GeV c$^{-1}$ verifies,
\begin{equation}
\label{eq::momentum}
    p^2 c^2 = E_0^2 - E_{\mu}^2 c^4 ,
\end{equation}
where $E_{\mu} = 0.10566$~GeV c$^{-2}$ is the muon mass and $c = 1$. $A_B$ , $a_0$ , $a_1$ , $a_2$, and $a_3$ are the fitting parameters, adjusted with different momentum ranges, as shown in Table \ref{tab::bugaev}.

\subsubsection{Reyna-Bugaev model}
Reyna et al.~\cite{Reyna2006} addressed the zenith angle independence of Bugaev's model by
\begin{equation}
    \Phi_R(p, \theta) = A_R \hat{p}^{-(a_3y^3+a_2y^2+a_1y+a_0)} \cos^3 \theta,
\end{equation}
where $y = \log_{10} \hat{p}$ and $\hat{p} = p \cos \theta$. The fitting parameters are $A_R = 0.00253$, $a_0 = 0.2455$, $a_1 = 1.288$, $a_2 = -0.2555$, and $a_3 = 0.0209$.
\begin{table*}
\begin{minipage}{106mm}
\caption{Bugaev's model parameters for the vertical energy spectrum of muons at sea level}
\label{tab::bugaev}
\begin{tabular}{cccccc}
\hline
$p$ range (GeV c$^{-1}$)  & $A_B$ (cm$^2$ sr s GeV)$^{-1}$ & $a_0$ & $a_1$  & $a_2$ & $a_3$   \\ \hline
1 - 930                     & 2.950 $\times 10^{-3}$    & 0.3061   & 1.2743     & -0.263       & 0.0252       \\
930 - 1590                  & 1.781 $\times 10^{-2}$      & 1.791     & 0.304      & 0   & 0    \\
1590 - 4.2$\times 10^5$     & 1.435 $\times 10^{1}$     & 3.672     & 0     & 0       & 0       \\
$>$ 4.2$\times 10^5$        & $10^{3}$     & 4.0     & 0      & 0   & 0    \\
\hline
\end{tabular}
\end{minipage}
\end{table*}

\subsubsection{Reyna-Hebbeker model}
Hebbeker $\&$ Timmermans \cite{Hebbeker2002} presented an empirical model similar to Bugaev's. It follows a power law depending on the muon momentum, 
\begin{equation}
    \Phi_H(p) = A_H 10^{H(y)},
\end{equation}
and
\begin{equation}
\begin{split}
    H(y) = h_1\frac{y^3 - 5y^2 + 6y}{2} + h_2 \frac{-2y^3 + 9y^2 - 10y + 3}{3} + \\
h_3 \frac{y^3 - 3y^2 + 2y}{6} + s_2 \frac{y^3 - 6y^2 + 11y - 6}{3},
\end{split}
\end{equation}
where $y = \log_{10} p$ with $p$ as defined by equation \ref{eq::momentum}. The fitting parameters are $A_H = 0.86$, $h_1 = 0.133$, $h_2 = -2.521$, $h_3 = -5.78$, and $s_2 = -2.11$ \cite{Lechmann2021}. Lesparre et al. \cite{Lesparre2012} introduce a zenith angle dependency to the model, obtaining: 
\begin{equation}
    \Phi_{RH}(p, \theta) = A_H 10^{H(\hat{y})} \cos^3 \theta,
\end{equation}
where $\hat{y} = \log_{10} (p \cos \theta)$.

\subsection{Altitude correction}
The semi-empirical models presented above estimate the muon flux at sea level. MUYSC corrects the altitude variation of the muon flux using the ratio,
\begin{equation}
    \frac{\phi(h)}{\phi(h=0)}= \exp(h/h_0),
\end{equation}
$h$ is the altitude of the observation point altitude (in meters), $h_0 = 4900 + 750 p$, and $p$ is the muon momentum in GeV \cite{Lesparre2012,Hebbeker2002}. 

\subsection{Muon energy loss}
\begin{figure}
\begin{center}
\includegraphics[width=0.46\textwidth]{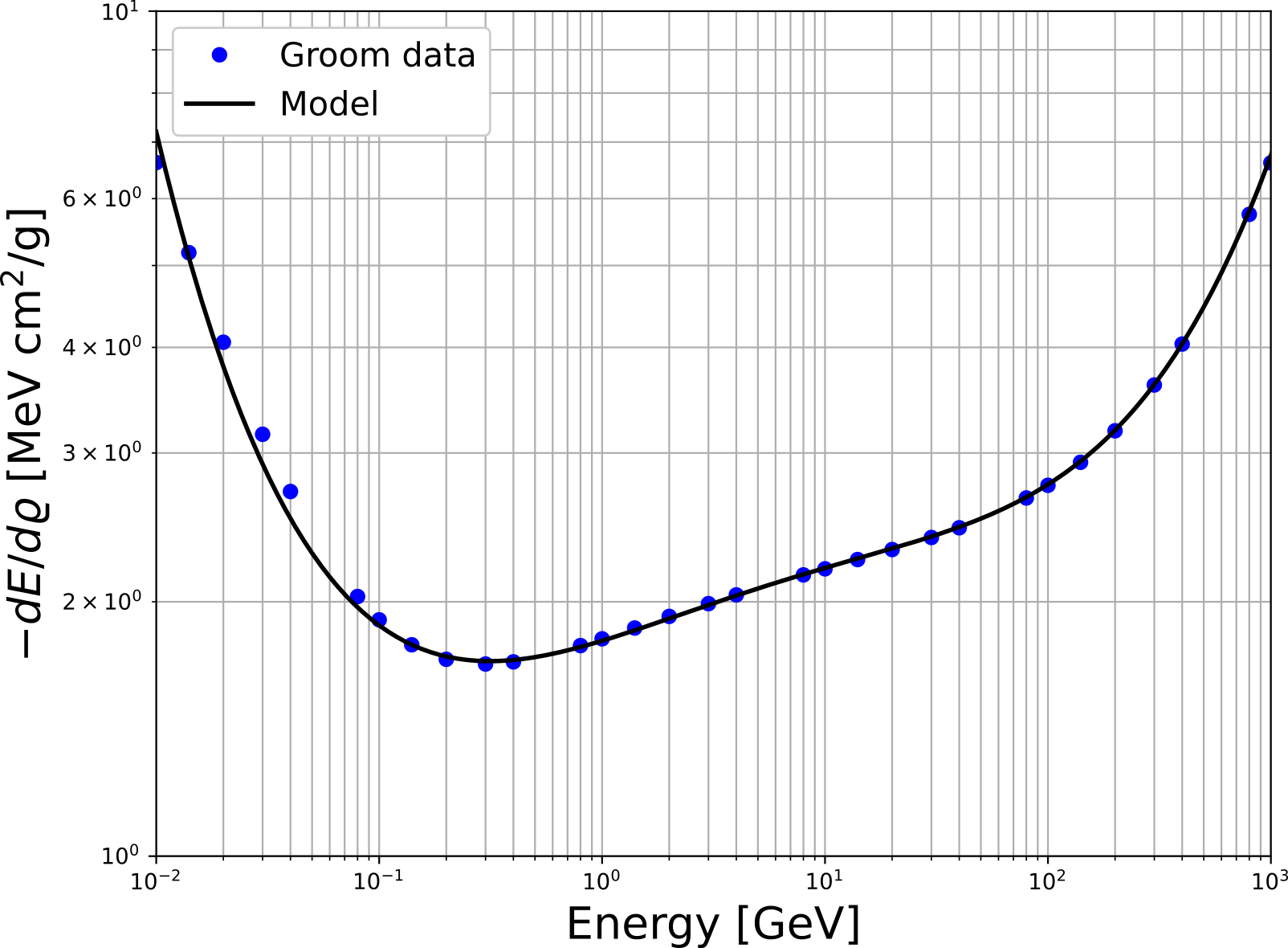}
\includegraphics[width=0.46\textwidth]{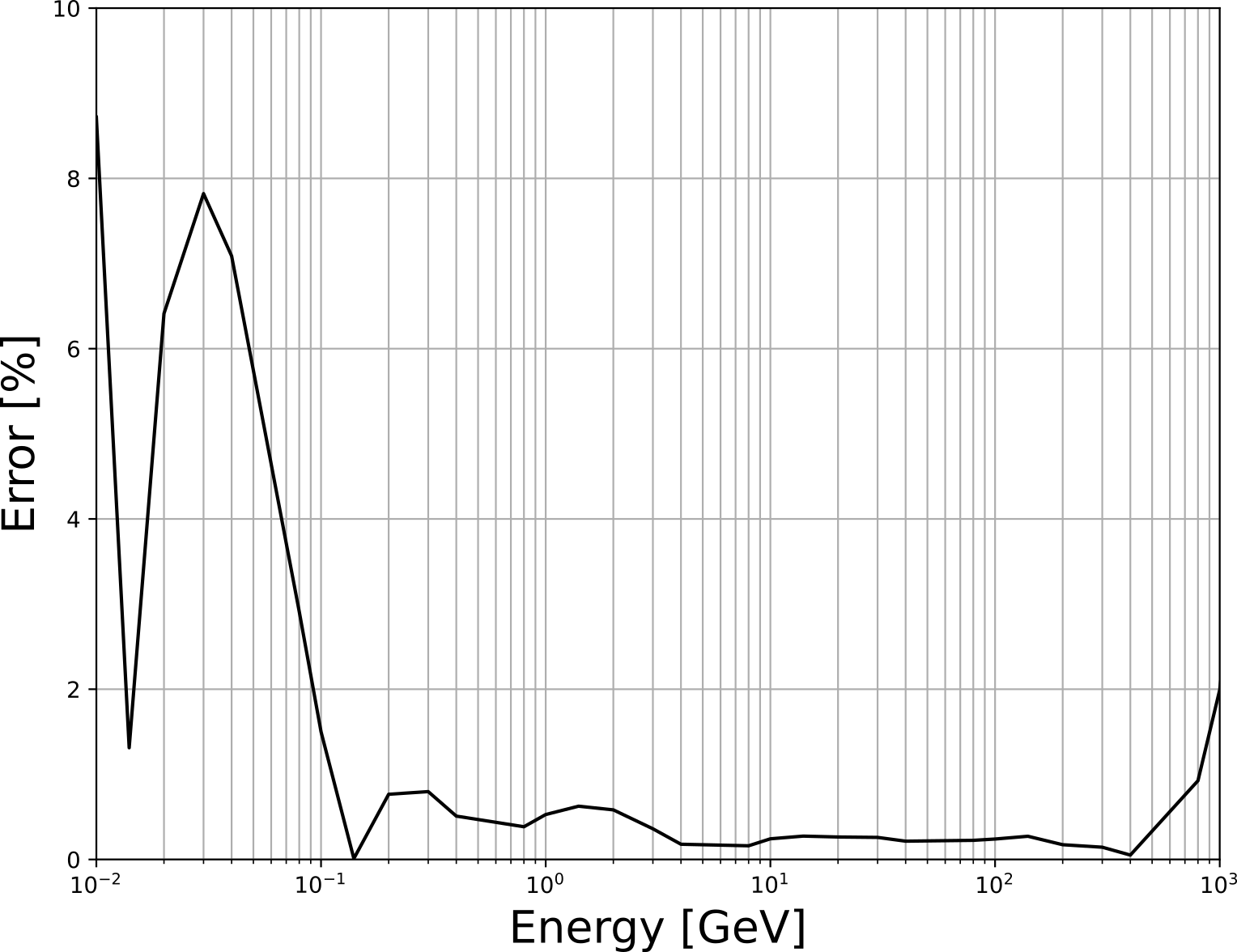}
\caption{Muon energy loss in standard rock. Lesparre's model (solid-line) fits the Groom's dataset (blue-points) with a maximum error of 8$\%$ for low energy muons ($< 10^{-1}$~GeV) and an average error $< 2\%$ for energies $\geq 10^{-1}$~GeV .}
\label{fig::energyloss}
\end{center}
\end{figure}

The energy loss is modelled as
\begin{equation}
    -\frac{dE}{d \varrho} = a(E) + b(E)E,
\end{equation}
where $a$ and $b$ are functions depending on the material properties through which the muons propagate and $\varrho(L)$ is the density integrated along the trajectory. Lesparre et al.~\cite{Lesparre2010} proposed the following model for the muon energy loss in standard rock based on the Groom dataset~\cite{Groom2001},
\begin{equation}
    \frac{dE}{d\varrho} = -10^{I_4x^4 + I_3x^3 + I_2x^2 +I_1x + I_0}.
\end{equation}
Notice that $x = \log E$, with $E$ the muon energy in GeV and $l_4 = 0.0154$, $l_3 = -0.0461$, $l_2 = 0.0368$, $l_1 = 0.0801$, $l_0 = 0.2549$. Fig. \ref{fig::energyloss} shows the performance of the muon energy loss model for energies from $10^{-2}$~GeV to $10^{3}$~GeV.

\subsection{Minimum energy estimation}
\begin{figure}
\begin{center}
\includegraphics[width=0.46\textwidth]{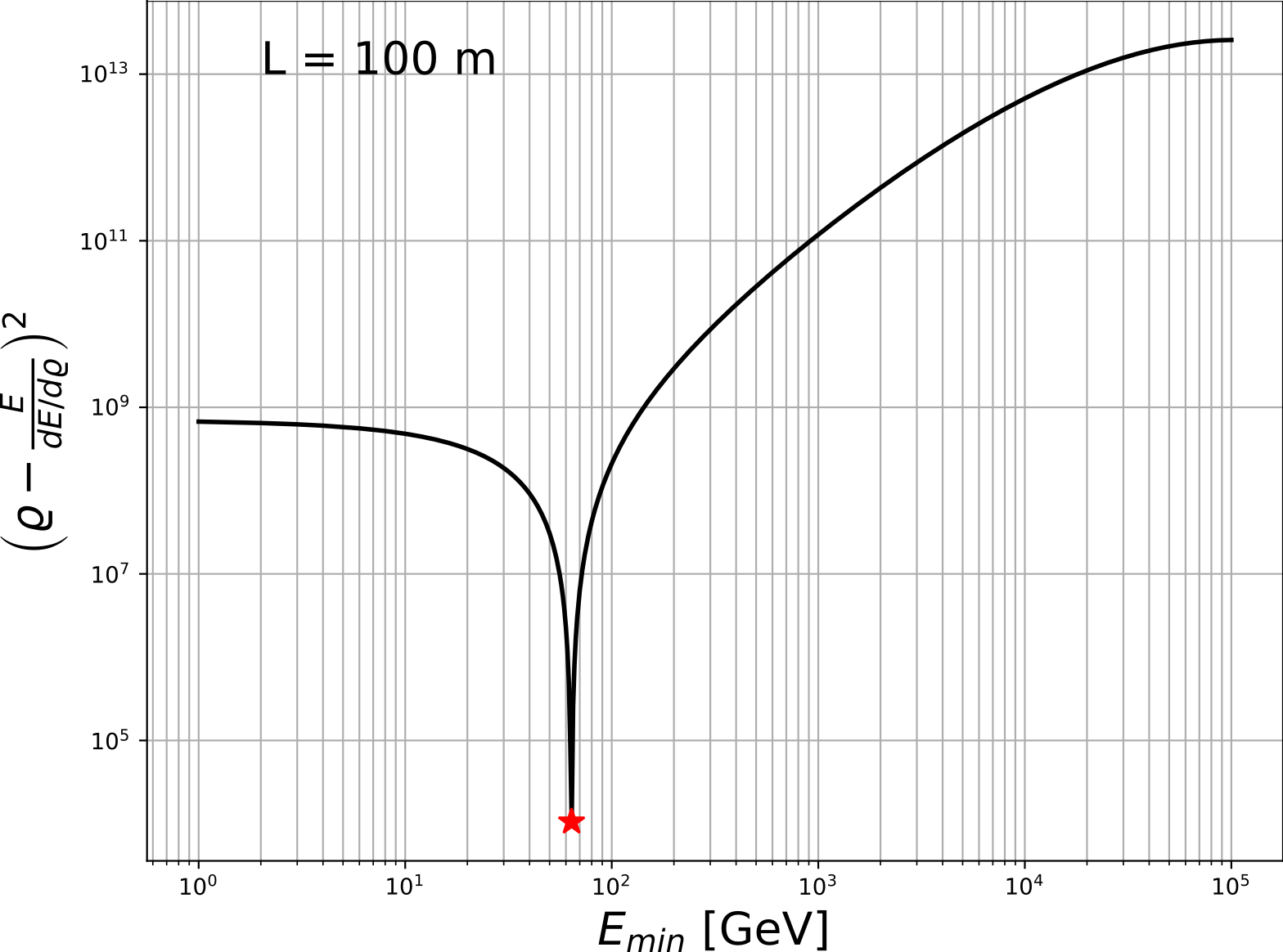}
\caption{MUYSC estimation of the minimum muon energy for traversing 100\,m of standard rock employing the minimization of the opacity error function. The minimum muon energy was $\sim 68.97$~GeV. }
\label{fig::minenergy}
\end{center}
\end{figure}

 The minimum muon energy for traversing a given amount of rock is,
\begin{equation}
    E_{min} = \int_0^{\varrho} \frac{dE}{d\varrho}d\varrho + E_{\mu},
\end{equation}
where $E_{\mu}$ is the muon mass and $\varrho$ the target opacity. The opacity integrates the material density along the muon path. MUYSC estimates the muon minimum energy, $E_{min} = E(a)$, by solving the following minimization problem
\begin{equation}
    a = \arg \min \left( \hat{\varrho} - \frac{E}{dE/d\varrho}\right)^2,
\end{equation}
where $\hat{\varrho} = \bar{\rho} L $ is the target opacity with $\bar{\rho} = 2.65$~g cm$^{-3}$ as the rock average density along the muon path $L$. The ratio $\frac{E}{dE/d\varrho}$ is obtained from the fitting model. Figure \ref{fig::minenergy} shows the minimum energy a muon needs to cross 100\,m of standard rock.
\subsection{Traversing muon flux}
Once MUYSC gets the minimum muon energy, it estimates the traversing muon flux through the target. MUYSC integrates the differential muon flux in energy from $E_{min}$ to infinity for the muon zenith angle as follows,
\begin{equation}
    I[\varrho,\theta] = \int_{E_{min}}^{\infty} \Phi(E_0,\theta)dE_0 \ [cm^{-2}sr^{-1}s^{-1}],
\end{equation}
and the discrete version, 
\begin{equation}
I(\varrho, \theta) = \sum_{E_{min}}^{\infty} \Phi(E_0,\theta)\Delta E_0 \ [cm^{-2}sr^{-1}s^{-1}].
\end{equation}

\begin{figure}
\begin{center}
\includegraphics[width=0.48\textwidth]{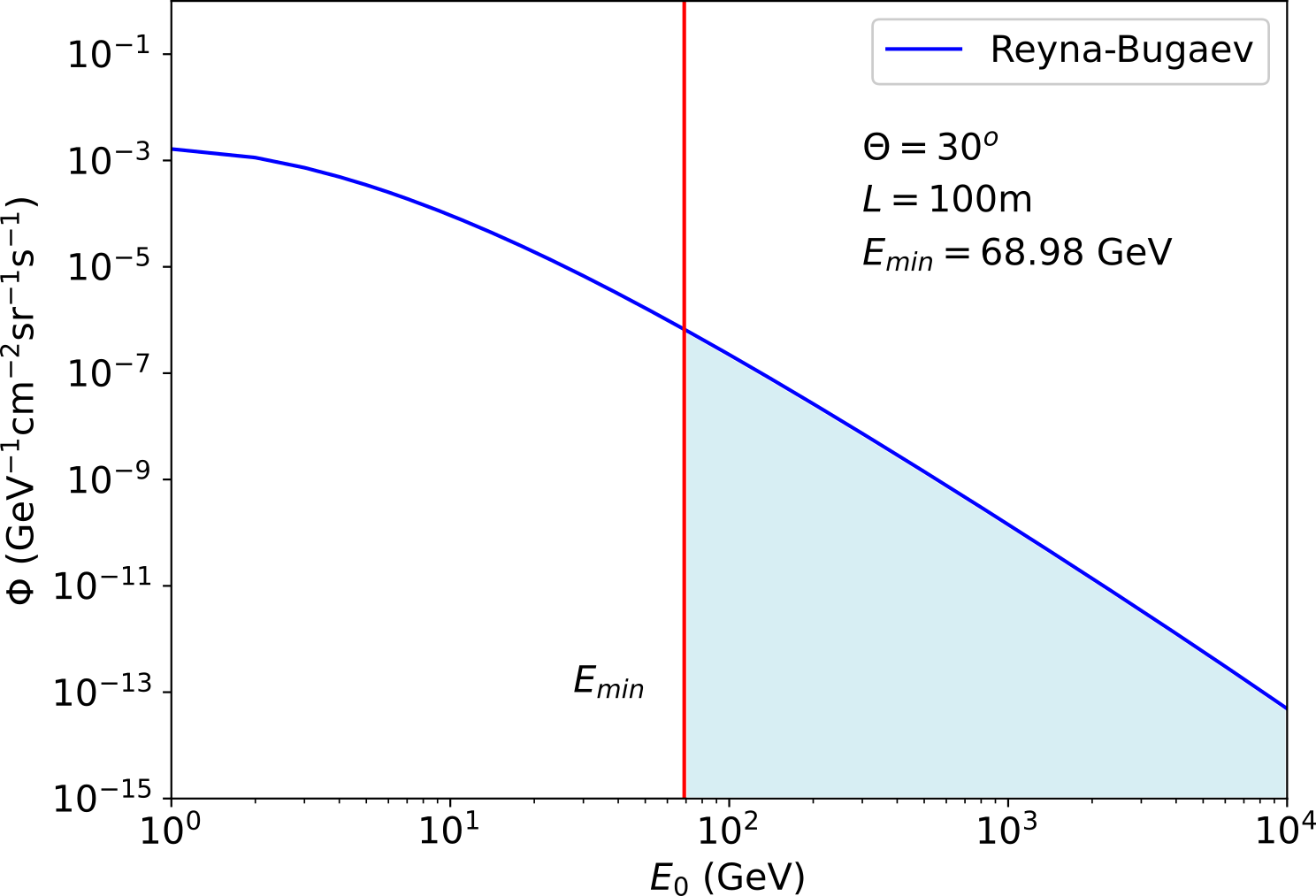}
\caption{Reyna-Bugaev differential muon flux for $\theta = 30^{\circ}$ (blue-line). The blue area represents the integrated muon flux passing along a rock thickness of $L = 100$~m. The muon's minimun energy is  $\sim 68.98$~GeV (red-line).}
\label{fig::integratedflux}
\end{center}
\end{figure}

The integrated muon flux $I(\varrho, \theta)$ depends on the differential muon flux $\Phi(E_0, \theta)$, the target opacity and the minimum muon energy. Figure \ref{fig::integratedflux} shows the differential flux integration region (light-blue) for a rock thickness of $L = 100\,m$ and a muon zenith angle of 30$^{\circ}$.

\subsection{Target density}
The integrated target density $\rho$ along the muon path $L$, leads to the opacity as 
\begin{equation}
    \varrho = \int_L \rho \ dl = - \frac{1}{\kappa} \ln{T} \ [\text{g cm}^{-2}],
\end{equation}
where $\kappa$ is the mass attenuation coefficient and $T$ the muon transmission \cite{PenarodriguezEng2021}. The ratio between the traversing $I$ and the open sky muon flux $I_0$ defines the muon transmission as follows,
\begin{equation}
    T = \frac{I}{I_0}.
\end{equation}
We get the target density by assuming an average density along the muon path
\begin{equation}
    \bar{\rho} = \frac{\varrho}{L}.
\end{equation}

\subsection{Validation of the MUYSC muon radiography module}
\begin{figure*}
\begin{center}
\vbox to220mm{\vfil
\xincludegraphics[width=0.4\textwidth ,label=\hspace{0.5cm}\color{white}{a)}, fontsize=\huge]{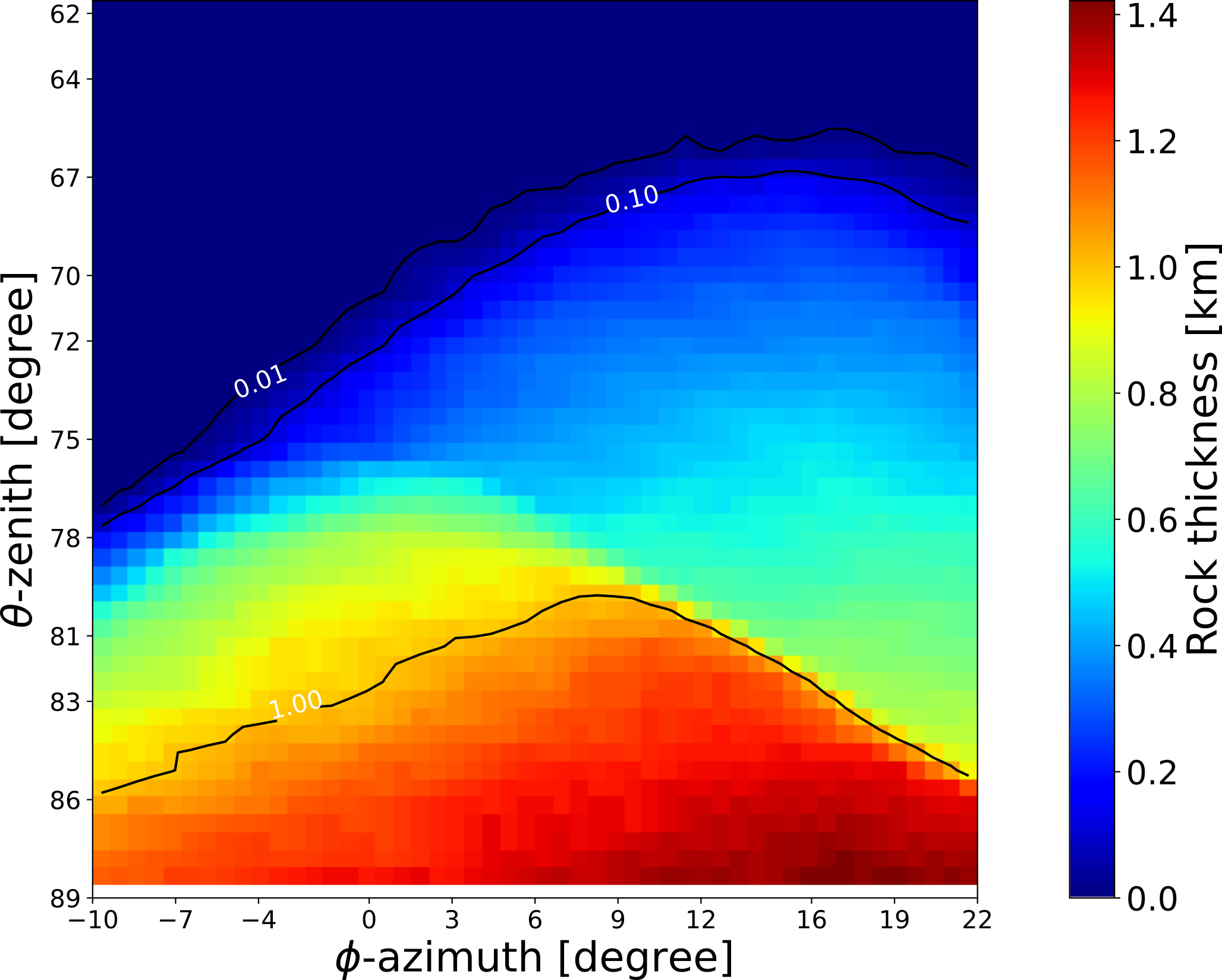}\hspace{0.5cm}
\xincludegraphics[width=0.4\textwidth ,label=\hspace{0.5cm}\color{white}{b)}, fontsize=\huge]{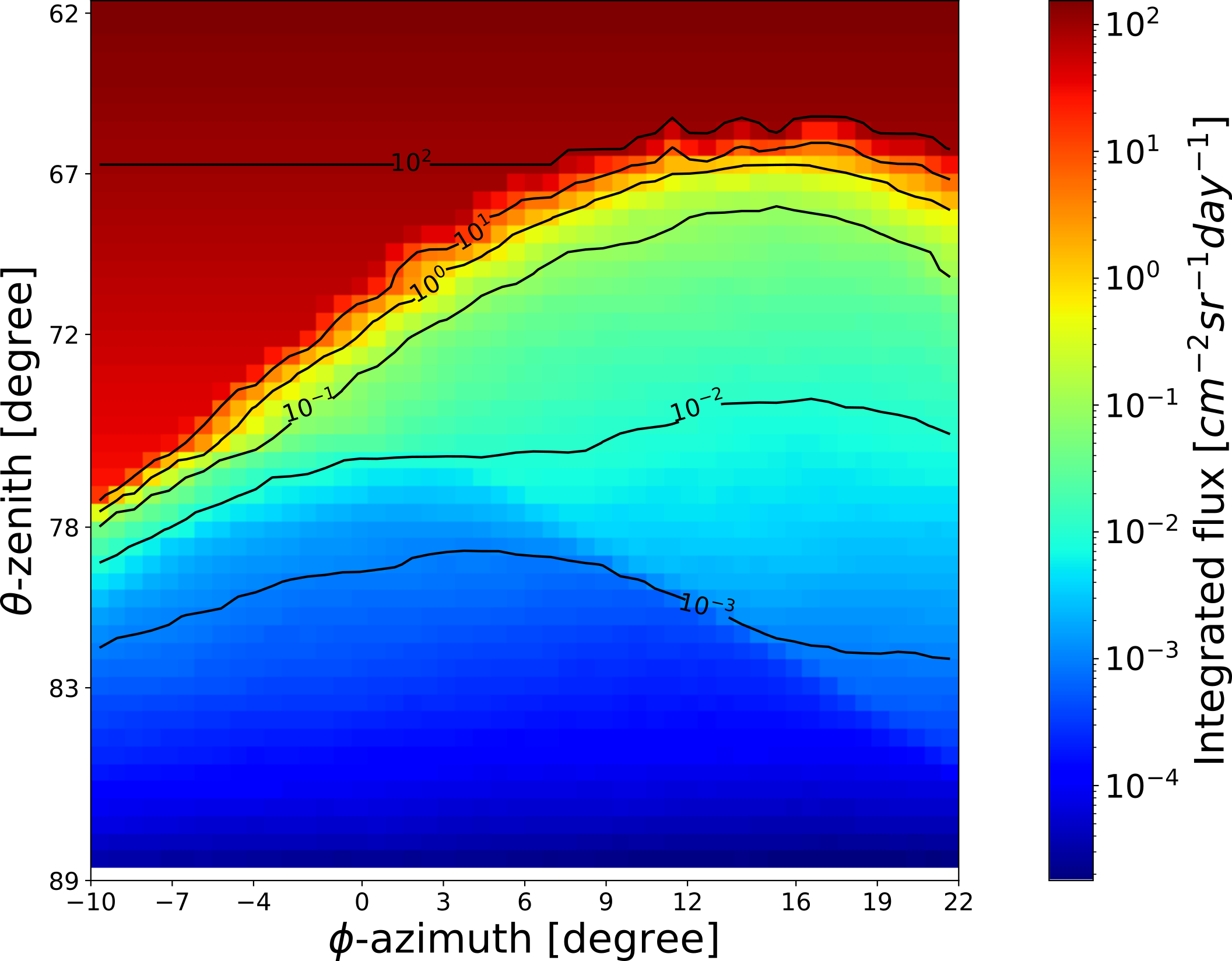}\\
\xincludegraphics[width=0.4\textwidth ,label=\hspace{0.5cm}\color{white}{c)}, fontsize=\huge]{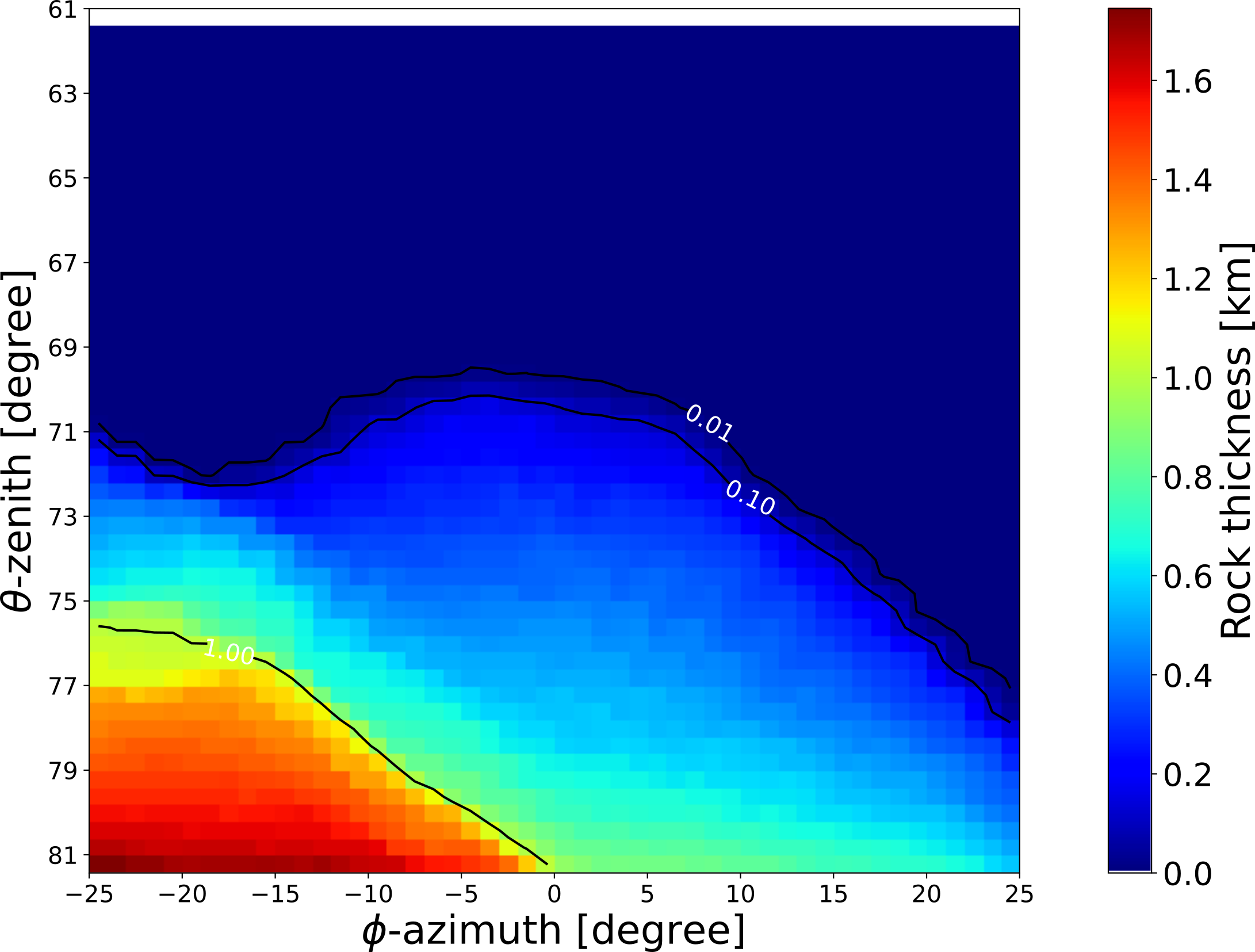} \hspace{0.5cm}
\xincludegraphics[width=0.4\textwidth ,label=\hspace{0.5cm}\color{white}{d)}, fontsize=\huge]{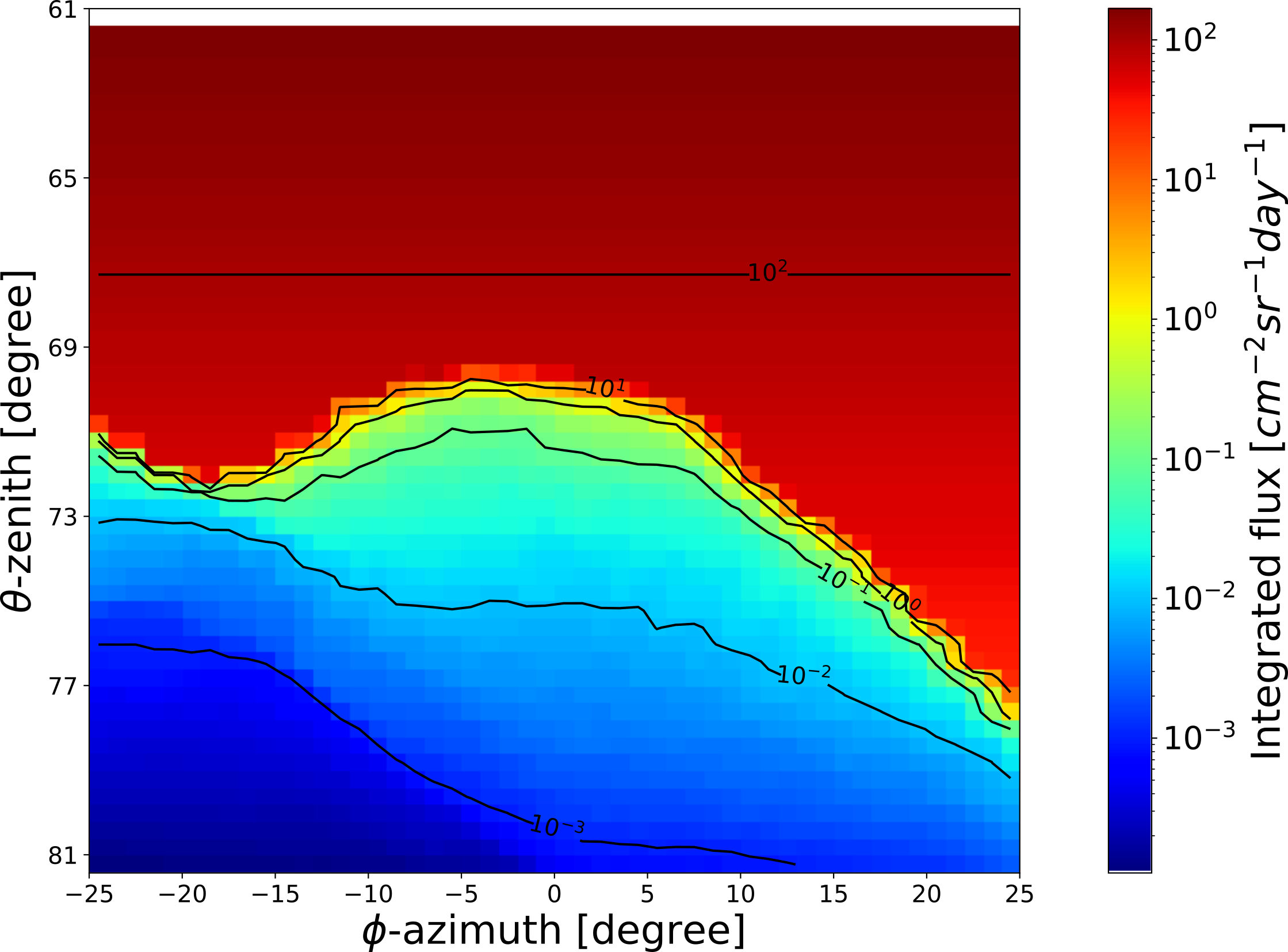} \\
\xincludegraphics[width=0.4\textwidth ,label=\hspace{0.5cm}\color{white}{e)}, fontsize=\huge]{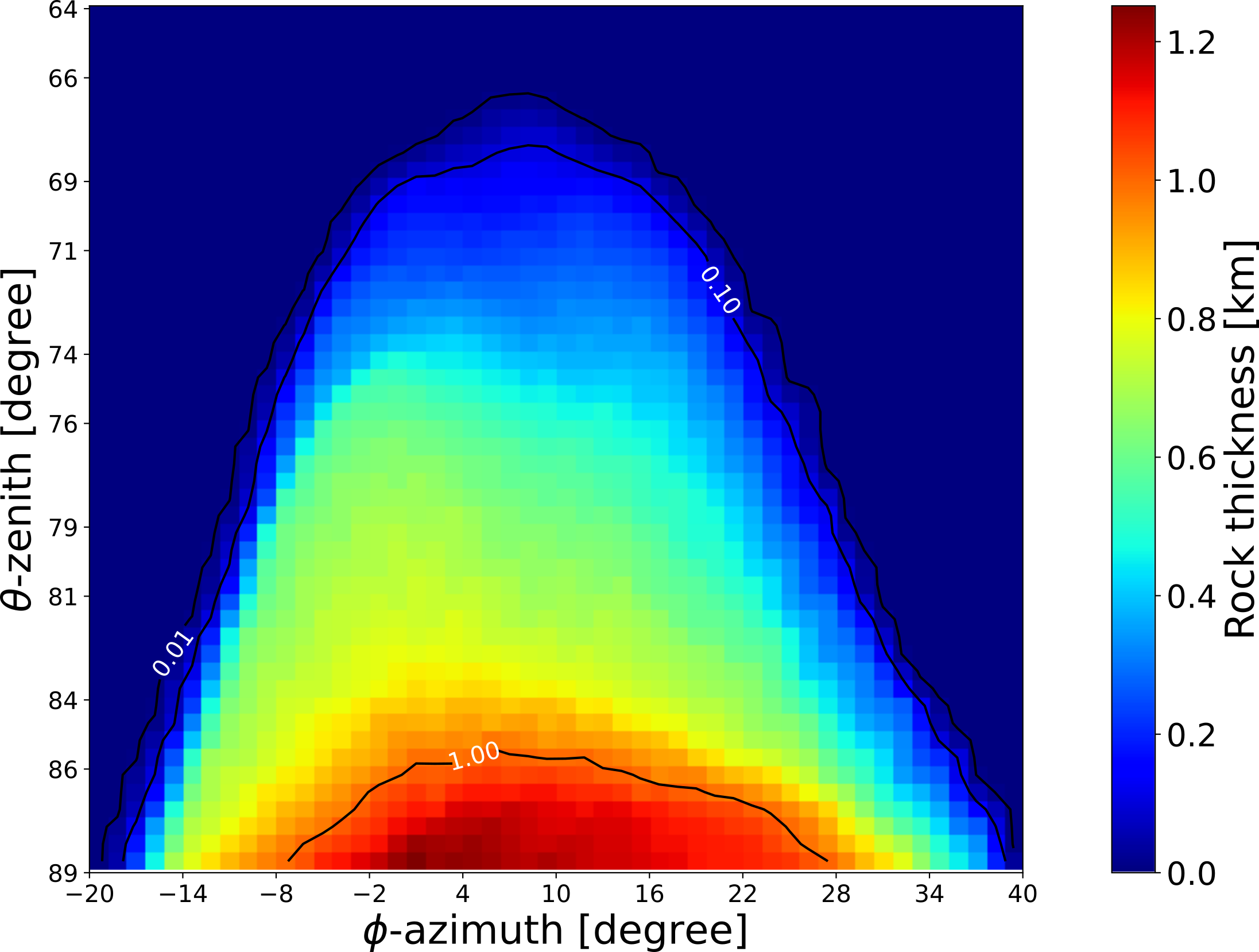} \hspace{0.5cm}
\xincludegraphics[width=0.4\textwidth ,label=\hspace{0.5cm}\color{white}{f)}, fontsize=\huge]{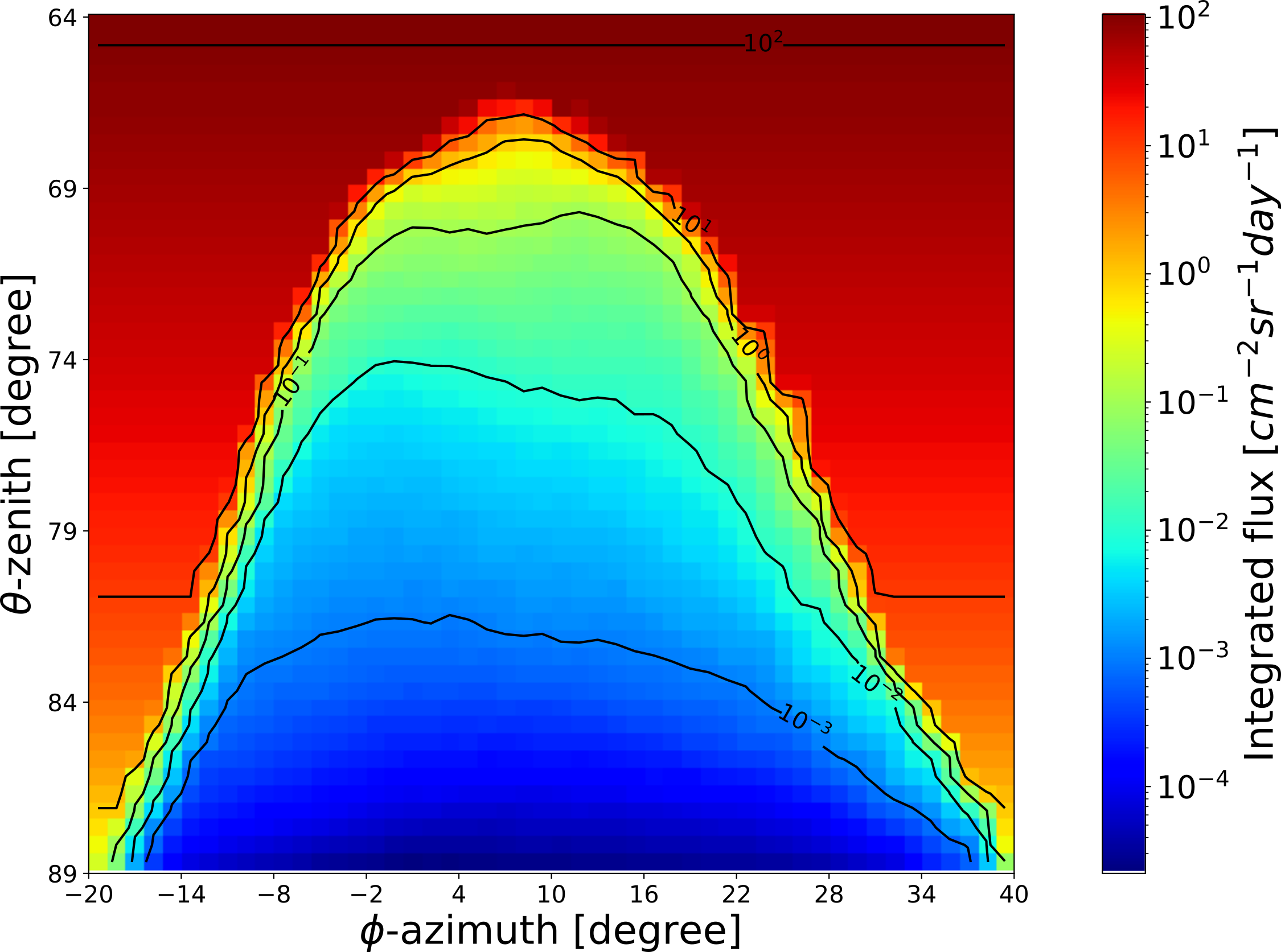}\\
\xincludegraphics[width=0.4\textwidth ,label=\hspace{0.5cm}\color{white}{g)}, fontsize=\huge]{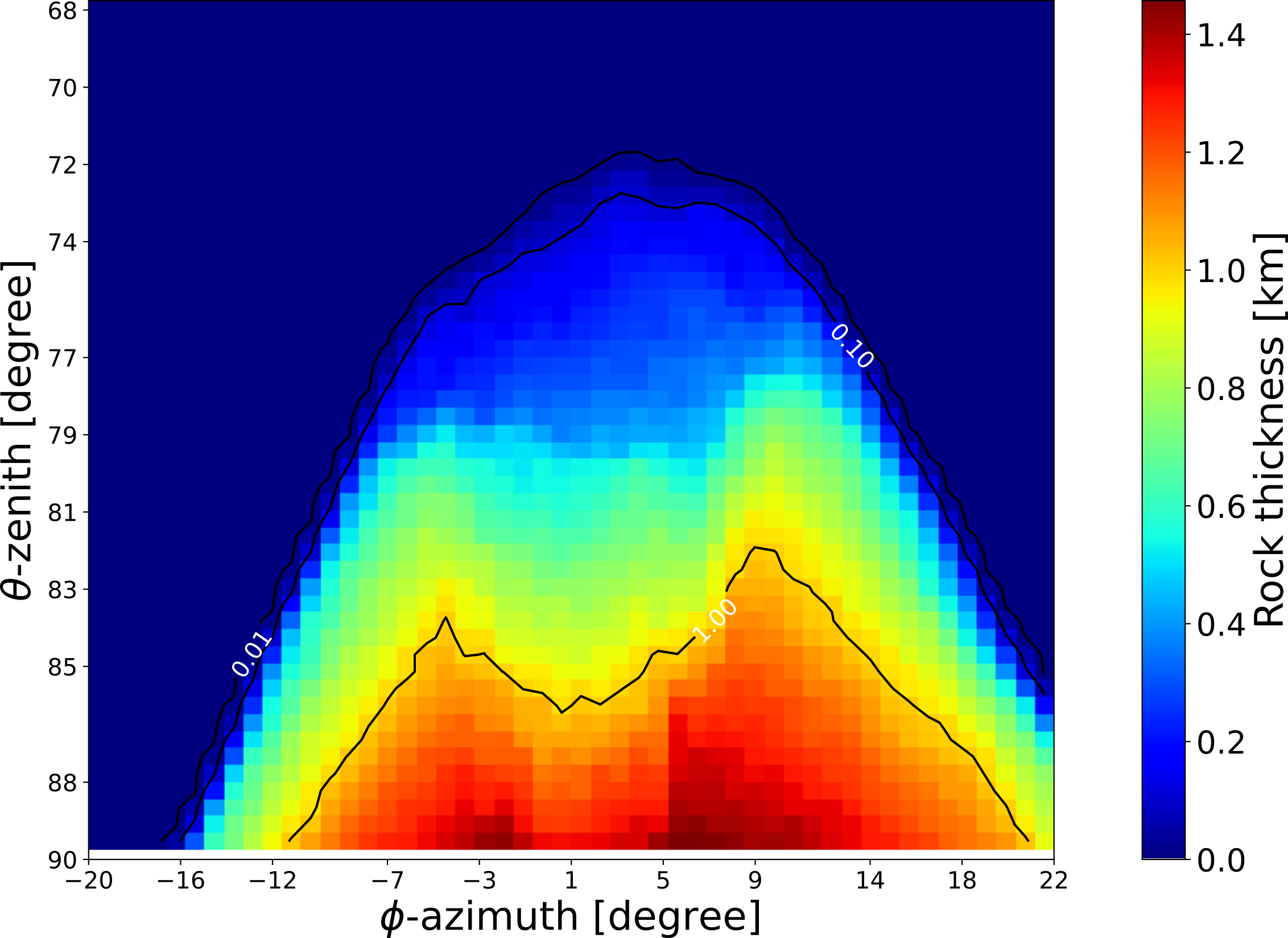} \hspace{0.5cm}
\xincludegraphics[width=0.4\textwidth ,label=\hspace{0.5cm}\color{white}{h)}, fontsize=\huge]{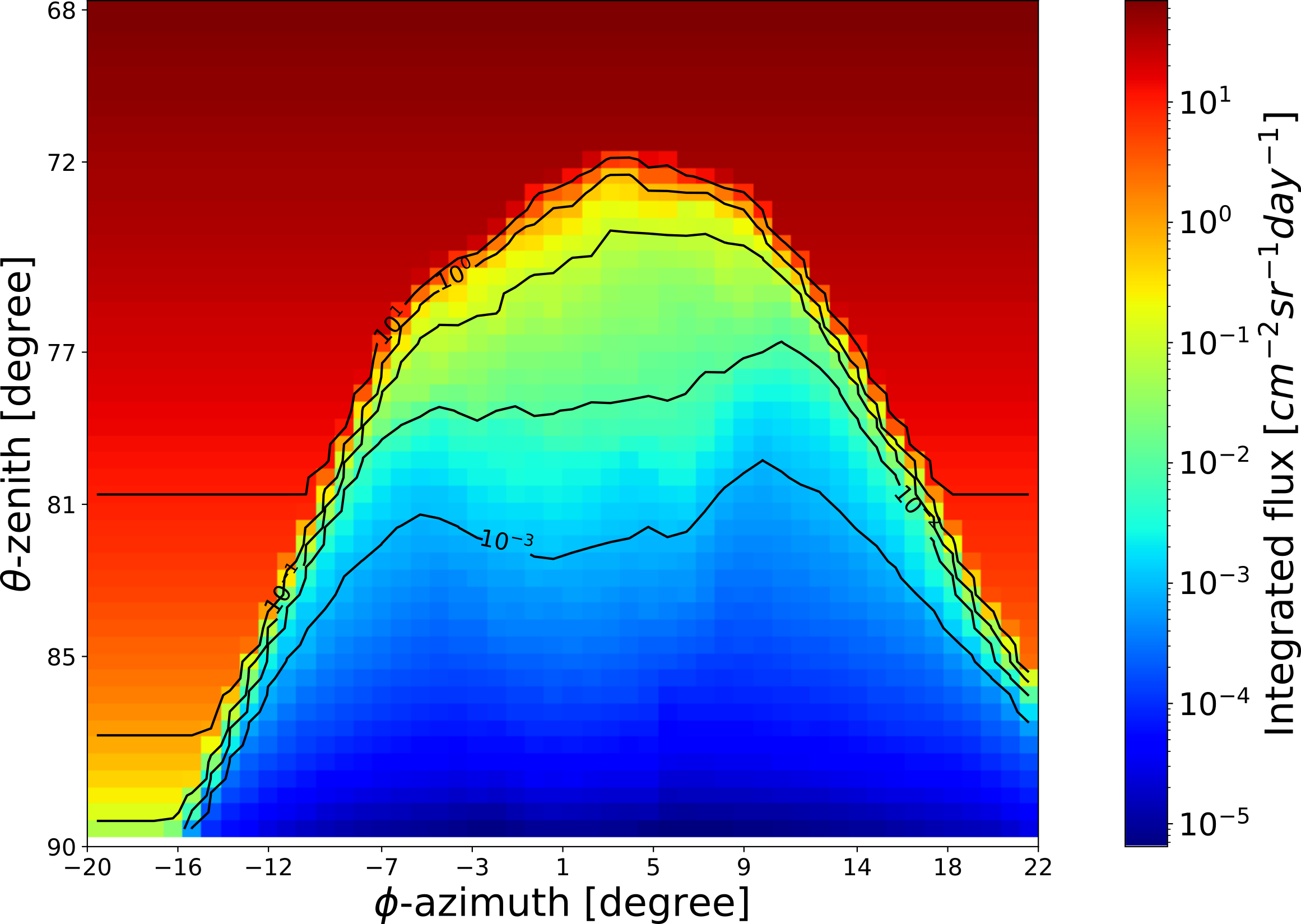}
\vfil}
\caption{Estimated rock thickness (a) and integrated muon flux (b) of the Cerro Machin volcano from the P1 observation point (4.492298$^{\circ}$ Latitude, -75.381092$^{\circ}$ Longitude). Estimated rock thickness (c) and integrated muon flux (d) of Mount Etna from the base observation point (14.996547$^{\circ}$ Latitude, 37.742343$^{\circ}$ Longitude). Estimated rock thickness (e) and integrated muon flux (f) of the Puy de Dôme from the Col de Ceyssat observation point (45.764167$^{\circ}$ Latitude, 2.955389$^{\circ}$ Longitude). Estimated rock thickness (gt) and integrated muon flux (h) of Mount Vesuvius from the Casina Amelia observation point (40.810561$^{\circ}$ Latitude, 14.410686$^{\circ}$ Longitude).}
\label{fig::mounts1}
\end{center}
\end{figure*}

We tested the MUYSC's muon radiography module by comparing the target rock thickness and the traversing integrated muon flux estimated by other experiments. MUYSC reproduced the results obtained with MuTe at the Cerro Machín volcano for the observation point P4 (4.494946 Latitude and -75.388110 Longitude) with an angular aperture of 30$^{\circ}$ azimuth and 25$^{\circ}$ zenith \cite{VesgaRamirez2020}. The maximum rock thickness (1.3\,km) occurs along the intersection of both volcano domes ($\theta = 84^{\circ}$, $\phi = 16^{\circ}$ to $19^{\circ}$), and the muon flux at the top of the dome reaching $10^2$ [cm$^{-2}$sr$^{-1}$ day$^{-1}$] as shown Fig. \ref{fig::mounts1}(b).

\begin{figure*}
\begin{center}
\vbox to60mm{\vfil
\includegraphics[width=0.4\textwidth]{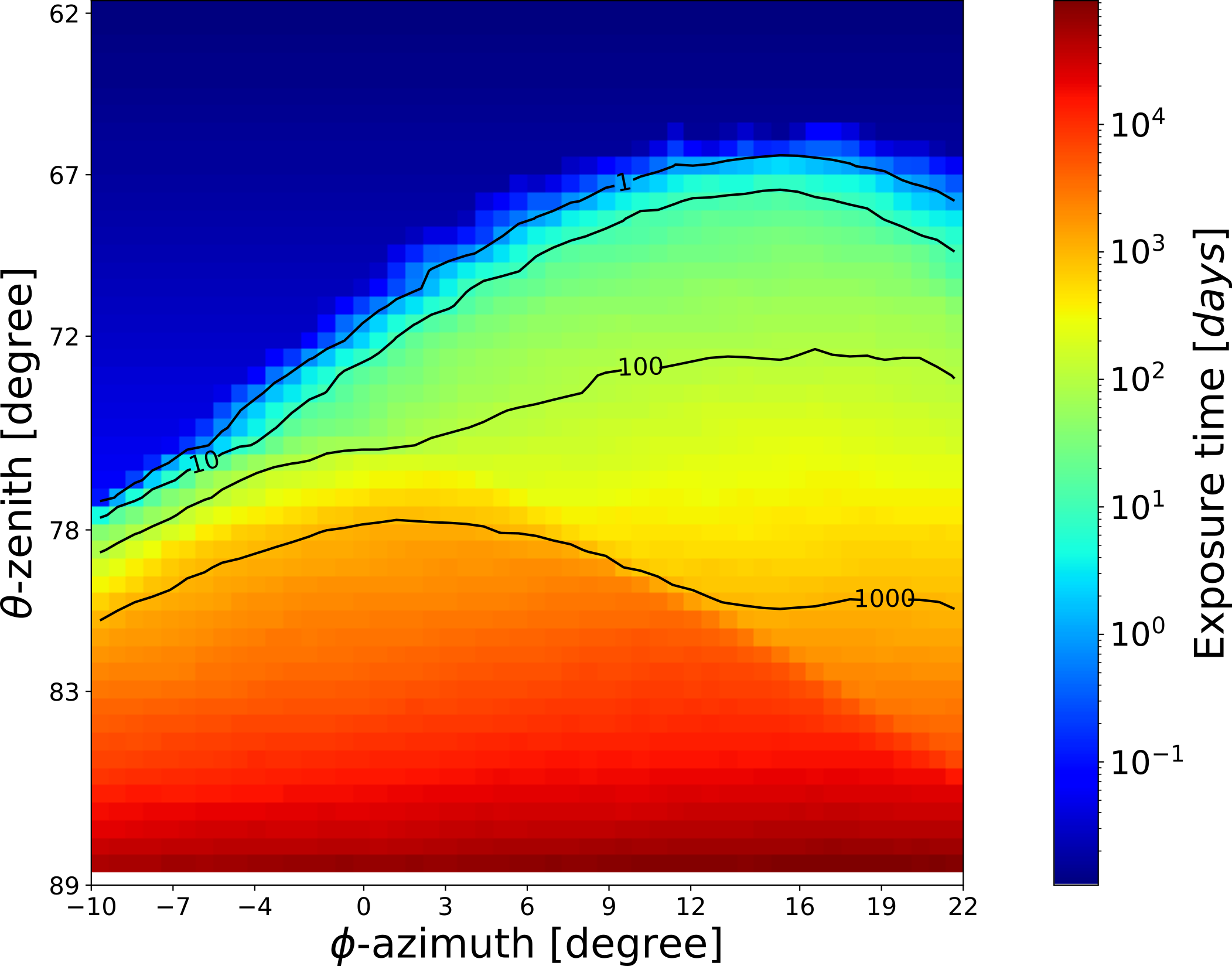} \hspace{0.5cm}
\includegraphics[width=0.4\textwidth]{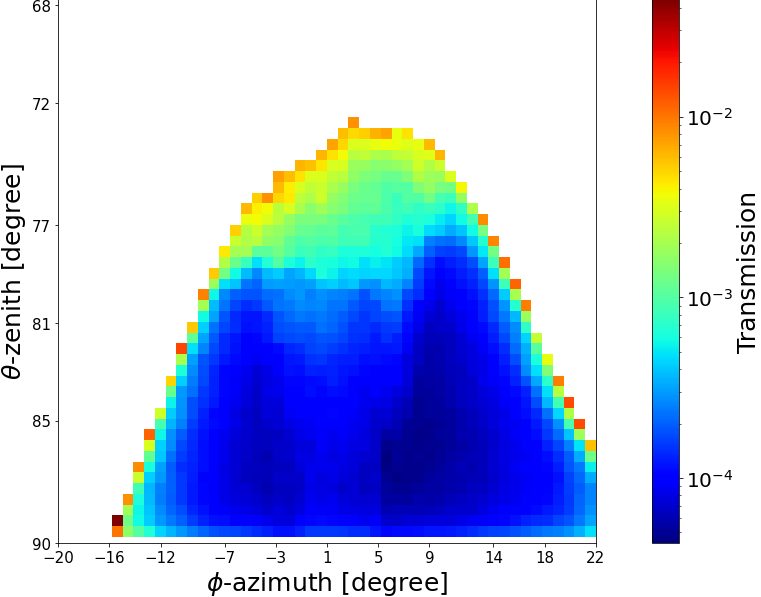}
\vfil}
\caption{(left) Expected exposure time of the Cerro Machin volcano from the P1 observation point (4.492298$^{\circ}$ Latitude, -75.381092$^{\circ}$ Longitude) for an acceptance of 6\,$cm^2sr^{-1}$ and a threshold of 10 muon/pixel. Expected muon transmission of Mount Vesuvius. }
\label{fig::muyscextra}
\end{center}
\end{figure*}

We simulated the Mount Etna muography from the observation point (14.996547$^{\circ}$ Latitude, 37.742343$^{\circ}$ Longitude) with an angular aperture of 50$^{\circ}$ azimuth and 20$^{\circ}$ zenith. MUYSC estimated a maximum rock thickness $\sim 1.7$~km ($\theta = 82^{\circ}$, $\phi = -25^{\circ}$), and a muon flux of  $2-3 \times 10^1$ [cm$^{-2}$sr$^{-1}$ day$^{-1}$] in the mountain borderland as shown Fig. \ref{fig::mounts1}(d). The result coincides with the work of Carbone et al.~\cite{Carnone2013}, but MUYSC resolves muon radiography with a higher resolution.

MUYSC also reproduced results obtained by the Diaphane experiment at the Puy de Dôme volcano \cite{Menedeu2016}. MUYSC sets the muon detector at the Col de Ceyssat observation point (45.764167$^{\circ}$ Latitude, 2.955389$^{\circ}$ Longitude). We observed a maximum rock thickness of around 1.2\,km ($\theta = 89^{\circ}$, $\phi = 3^{\circ}$), and a muon flux of $2-3 \times 10^1$ [cm$^{-2}$sr$^{-1}$ day$^{-1}$] at the volcano top as shown Fig. \ref{fig::mounts1}(f).

Fig. \ref{fig::mounts1}(g) shows how MUYSC can solve inner structures of volcanoes, i.e. craters. We simulated muon radiography of Mount Vesuvius from the Casina Amelia (40.810561$^{\circ}$ Latitude, 14.410686$^{\circ}$ Longitude) observation point with an angular aperture of 22$^{\circ}$ zenith and 42$^{\circ}$ azimuth pointing to the top of the volcano edifice. The maximum rock thickness (1.45\,km) around the base of the volcano crater ($\theta = 90^{\circ}$, $\phi = -3^{\circ}$ to $9^{\circ}$), and the muon flux arises to $1-2 \times 10^1$ [cm$^{-2}$sr$^{-1}$ day$^{-1}$] at the top of the volcano \cite{Enricco2022}.

MUYSC can also compute the target transmission matrix, the detector exposure time for a given muon rate per pixel threshold, and the target opacity assuming a homogeneous material along the muon path (e.g. standard rock density $\rho = 2.65 g/cm^3$). Fig. \ref{fig::muyscextra}-left shows the exposure time estimated by MUYSC for the muography of the Cerro Machin volcano from the P1 observation point with a uniform detector acceptance of 6\,$cm^2sr^{-1}$ and a threshold of 10 muon/pixel \cite{VesgaRamirez2020}. The exposure time in regions with low rock thickness ($\sim 10^1$~m) spans a couple of days, but in areas where rock thickness increases to 1\, km, the exposure time reaches 1000 days.

Figure \ref{fig::muyscextra}-(right) shows the transmission matrix of Mount Vesuvius. Muon transmission is the ratio between the traversing and open sky muon flux. The muon transmission reaches $5 \times 10^{-2}$ at the volcano border and decreases with the rock thickness till $1 \times 10^{-3}$ at the volcano crater. Such MUYSC results agree with the work of Alessandro et al.~\cite{Alessandro2018}.

\section{Detector parameterization}
\label{chap::telescope}
The detector parameterization module computes the angular resolution and acceptance of any muon telescope by inputting the pixel area, the number of pixels per sensitive matrix and the distance between them.
\begin{figure}
\centering
\includegraphics[width=0.2\textwidth]{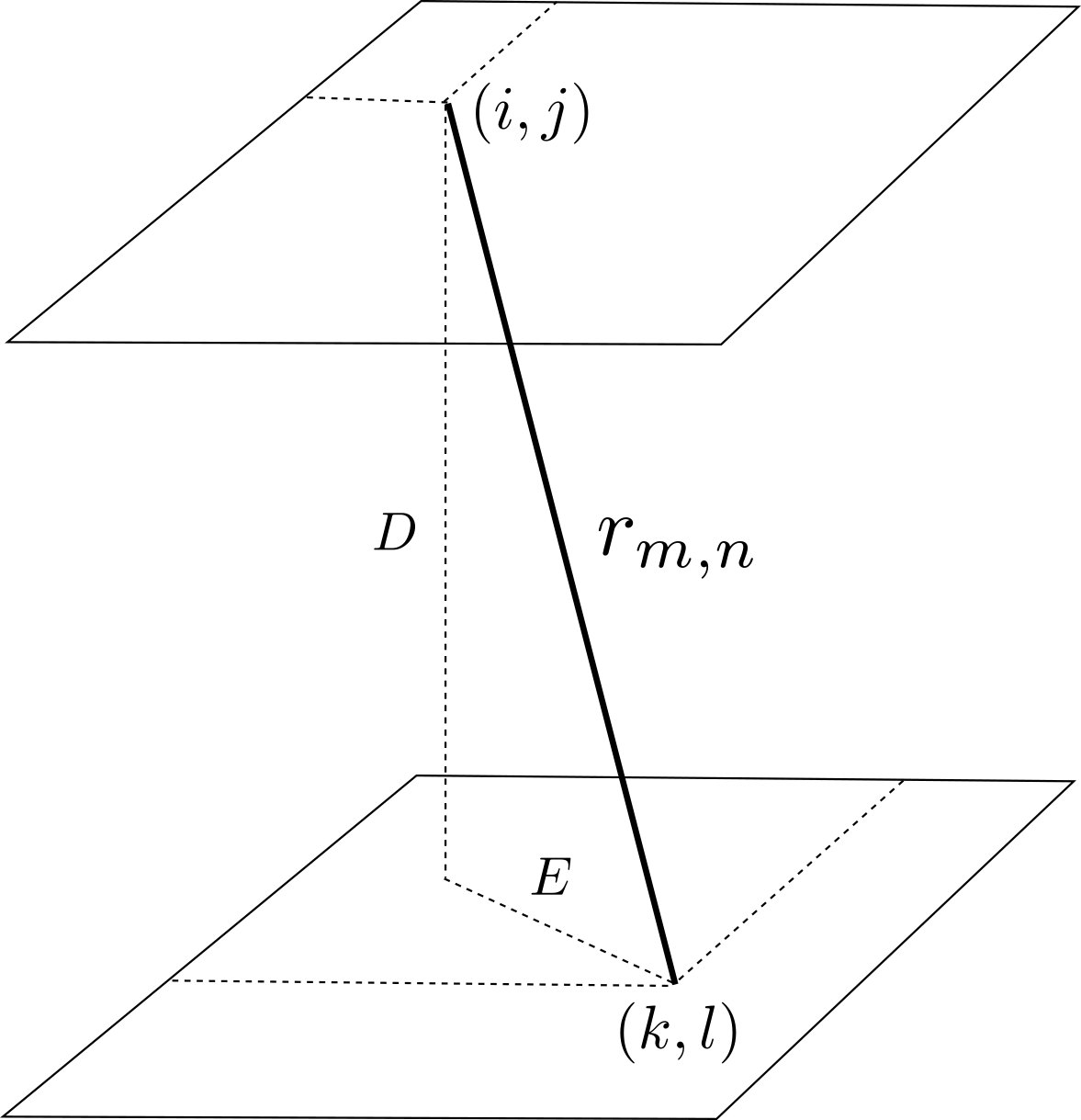}
\includegraphics[width=0.2\textwidth]{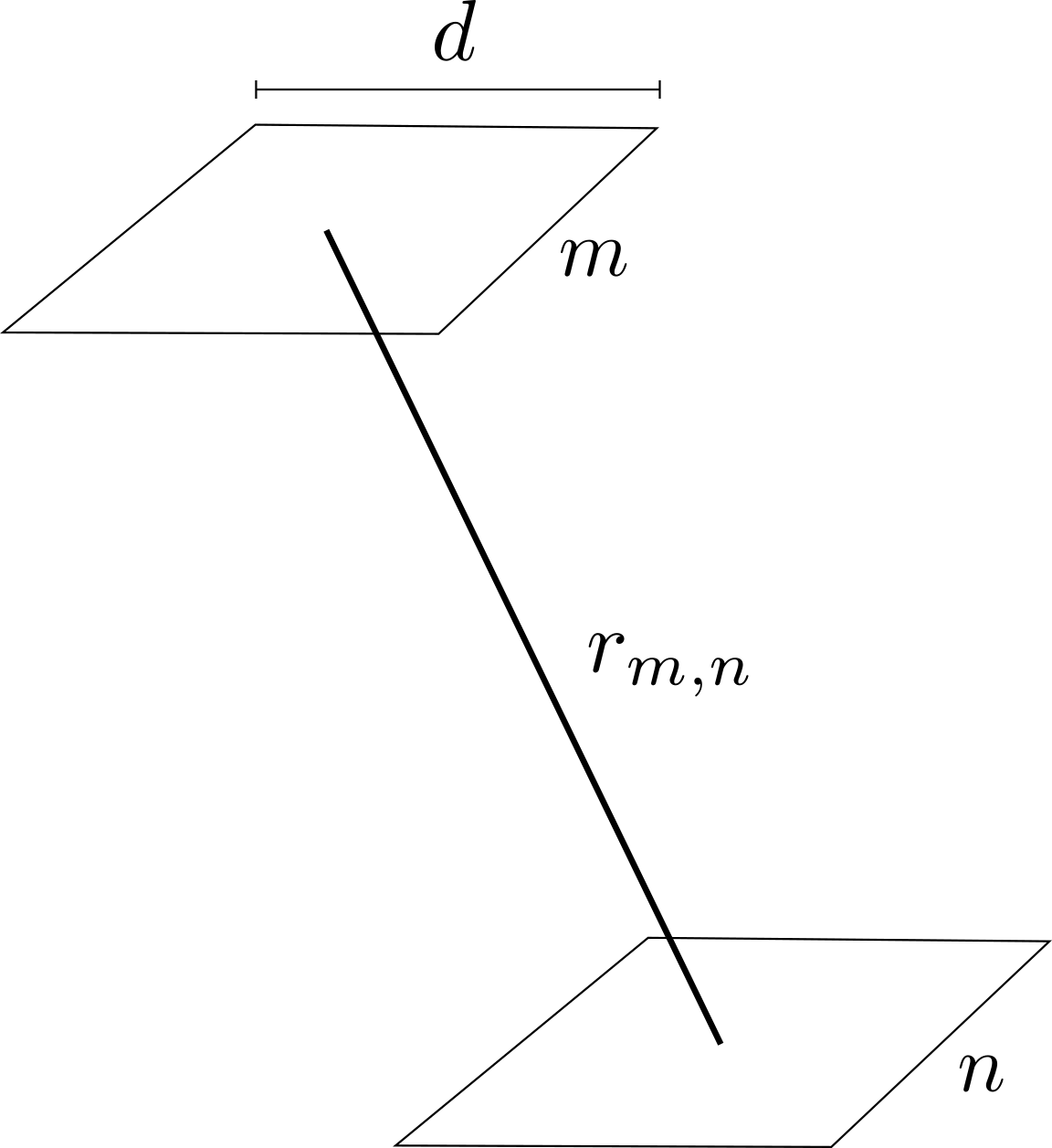}
\caption{Tracking of muons crossing the muon hodoscope. The activated pixels $m(i,j)$ and $n(k,l)$ defines the $r_{m,n}$ trajectory between the hodoscope layers. The trajectory distance depends on the separation between the layers $D$ and the relative distance between pixels $E$.}
\label{HodTraj}
\end{figure}

Telescope acceptance, $\mathcal{T}(r_{m,n})$, depends on the detection area, $S(r_{m,n})$, and solid angle, $\delta\Omega(r_{m,n})$)~\cite{Lesparre2010}, as,
\begin{equation}
\mathcal{T}(r_{m,n})=S(r_{m,n})\times \delta\Omega(r_{m,n}), \quad \text{where} \; \delta\Omega= \frac{4A}{r^2_{m,n}} \,, 
\end{equation}
with $A = d^2$ the pixel area, $d$ the pixel side, and $r_{m,n}$ the trajectory between the pixels $m$ and $n$. The trajectory depends on the distance between the detection panels $D$ and the relative distance $E$ between the pixels $m$ and $n$ as shown in Figure \ref{HodTraj}:
\begin{equation}
    r_{m,n}= \sqrt{D^2+E^2} \; \Leftrightarrow \; E= \sqrt{(| k-i | d)^2+ ( | l-j | d)^2 } \, , 
\end{equation}
where $i,j$ are the front panel $m$-pixel coordinates and $k,l$ the rear panel $n$-pixel coordinates.

The detection area depends on the number of panel-activated pixels, $N_P$, per trajectory, $r_{m,n}$, and the pixel area, $A$, as
\begin{equation}
S(r_{m,n}) = N_PA \, . 
\end{equation}
The muon hodoscope reconstructs $2N_{i-1}\times 2N_{j-1}$ trajectories, where $N_i$ is the number of $i$-bars and $N_j$ corresponds to $j$-bars. The traversing flux $I$ is defined as
\begin{equation}
I(r_{m,n}) = \frac{N(r_{m,n})}{\Delta T \times \mathcal{T}(r_{m,n})} \, ,
\end{equation}
where $N(r_{m,n})$ is the number of detected particles and $\Delta T$ is the detection time.

\begin{figure*}
\centering
\vbox to200mm{\vfil 
\includegraphics[width=.45\textwidth]{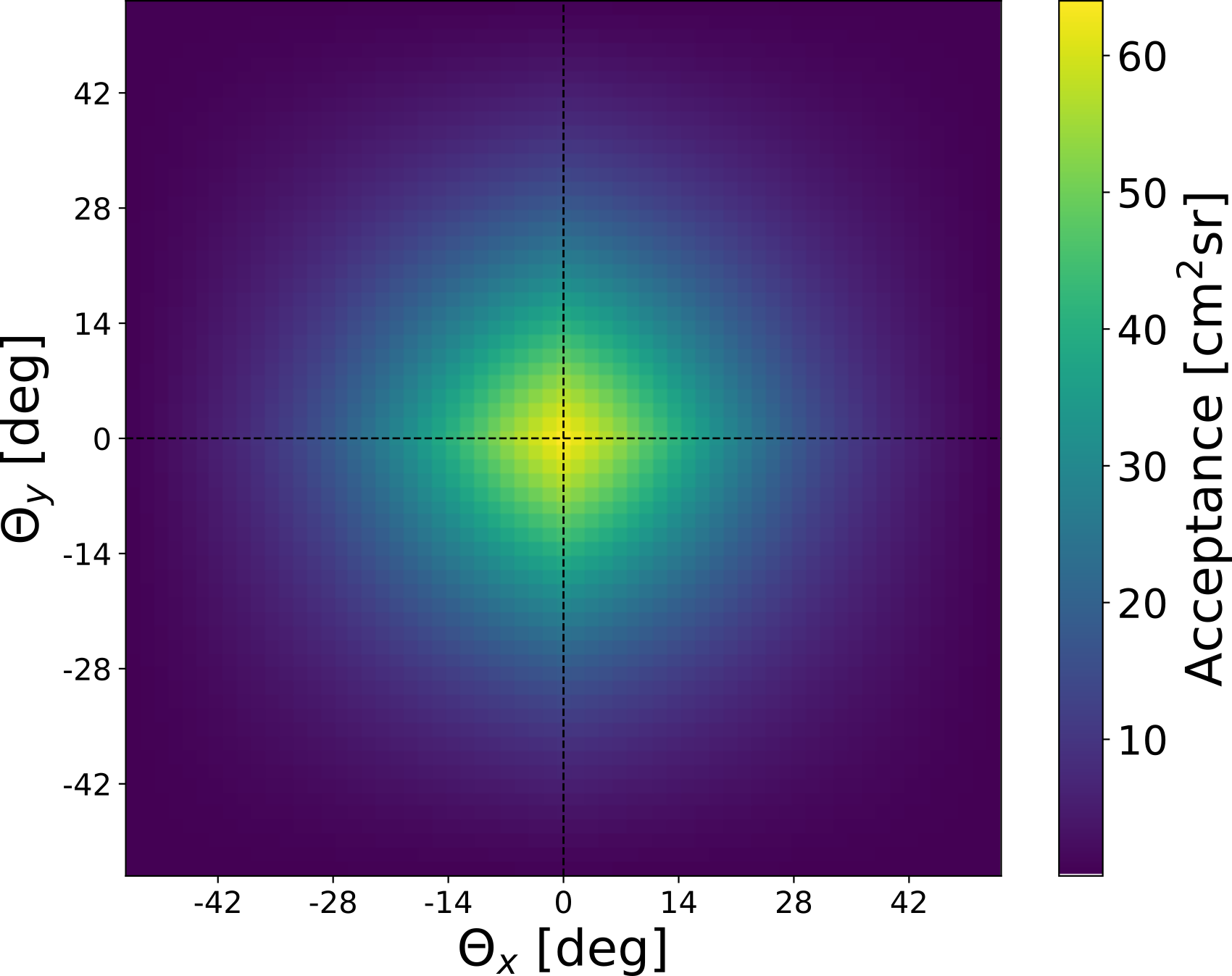}
\includegraphics[width=.45\textwidth]{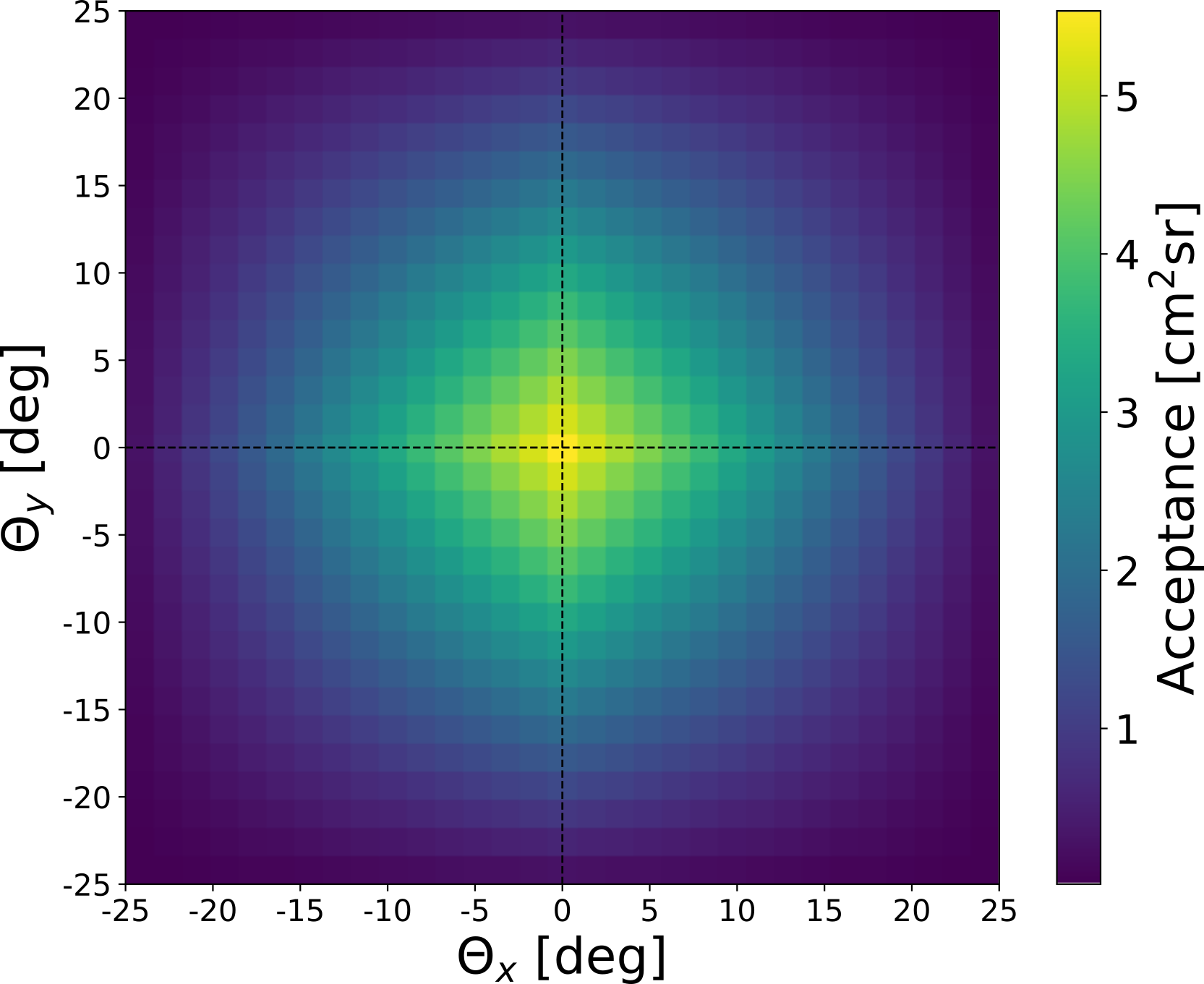} \\
\includegraphics[width=.45\textwidth]{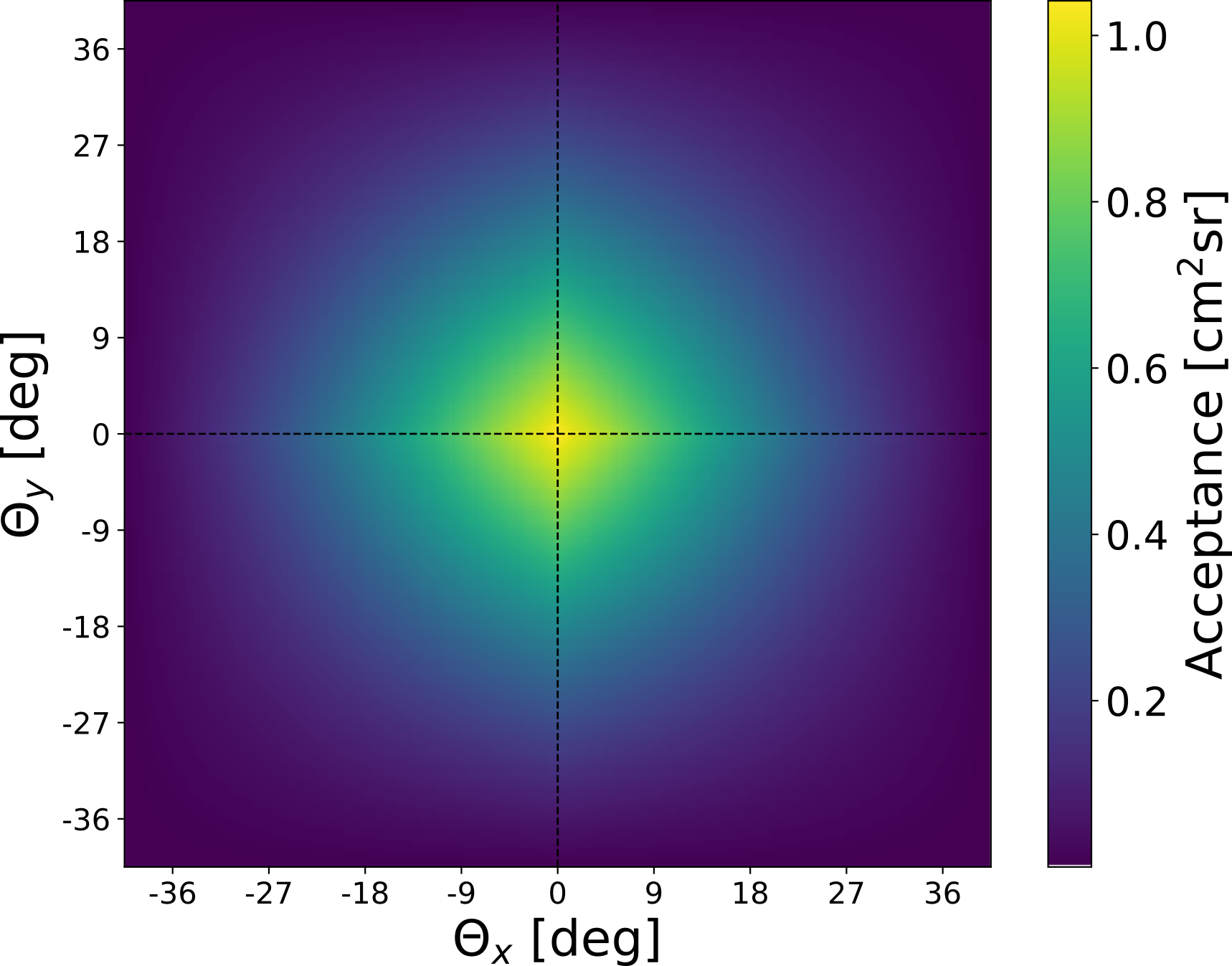}
\includegraphics[width=.45\textwidth]{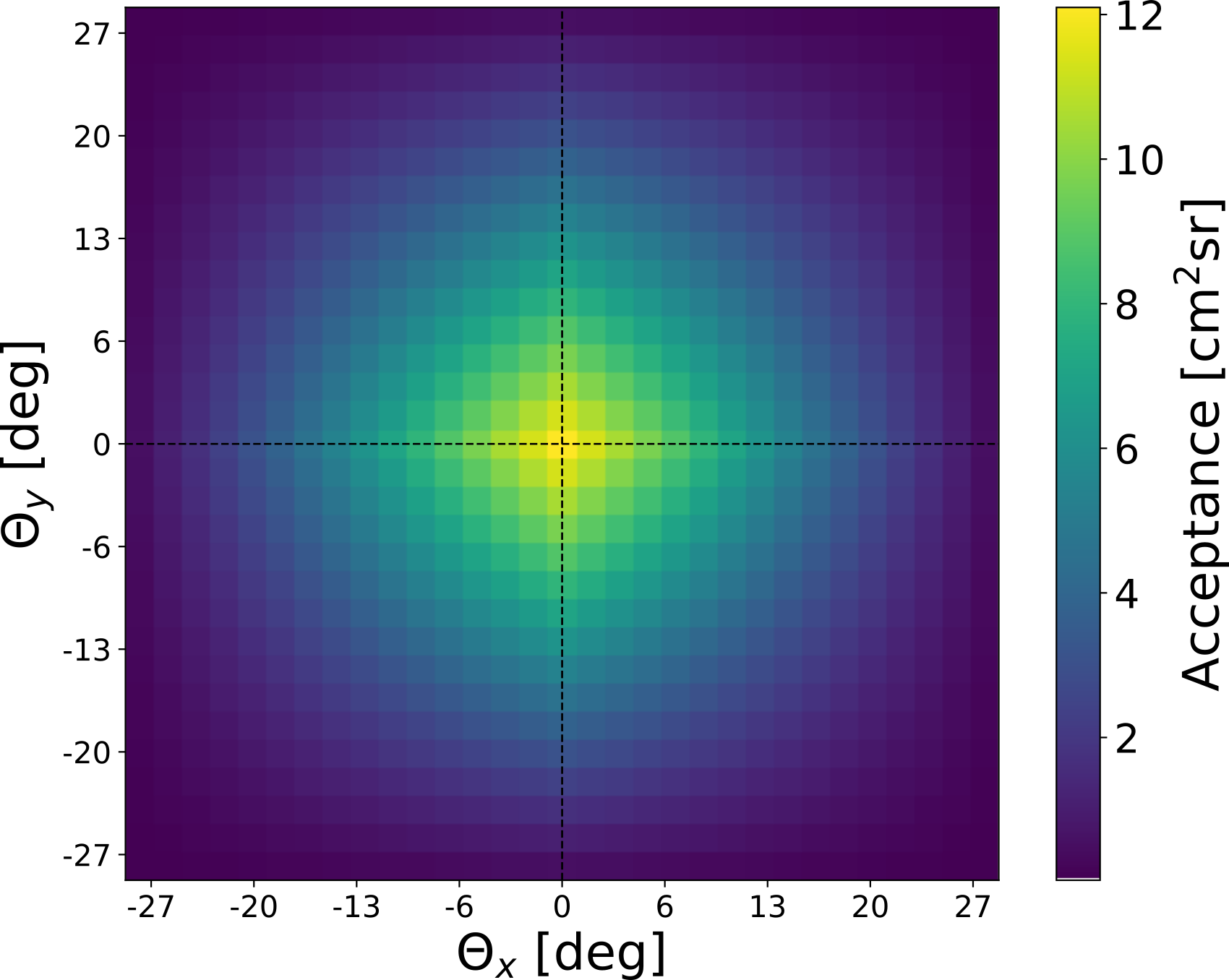} \\
\includegraphics[width=.45\textwidth]{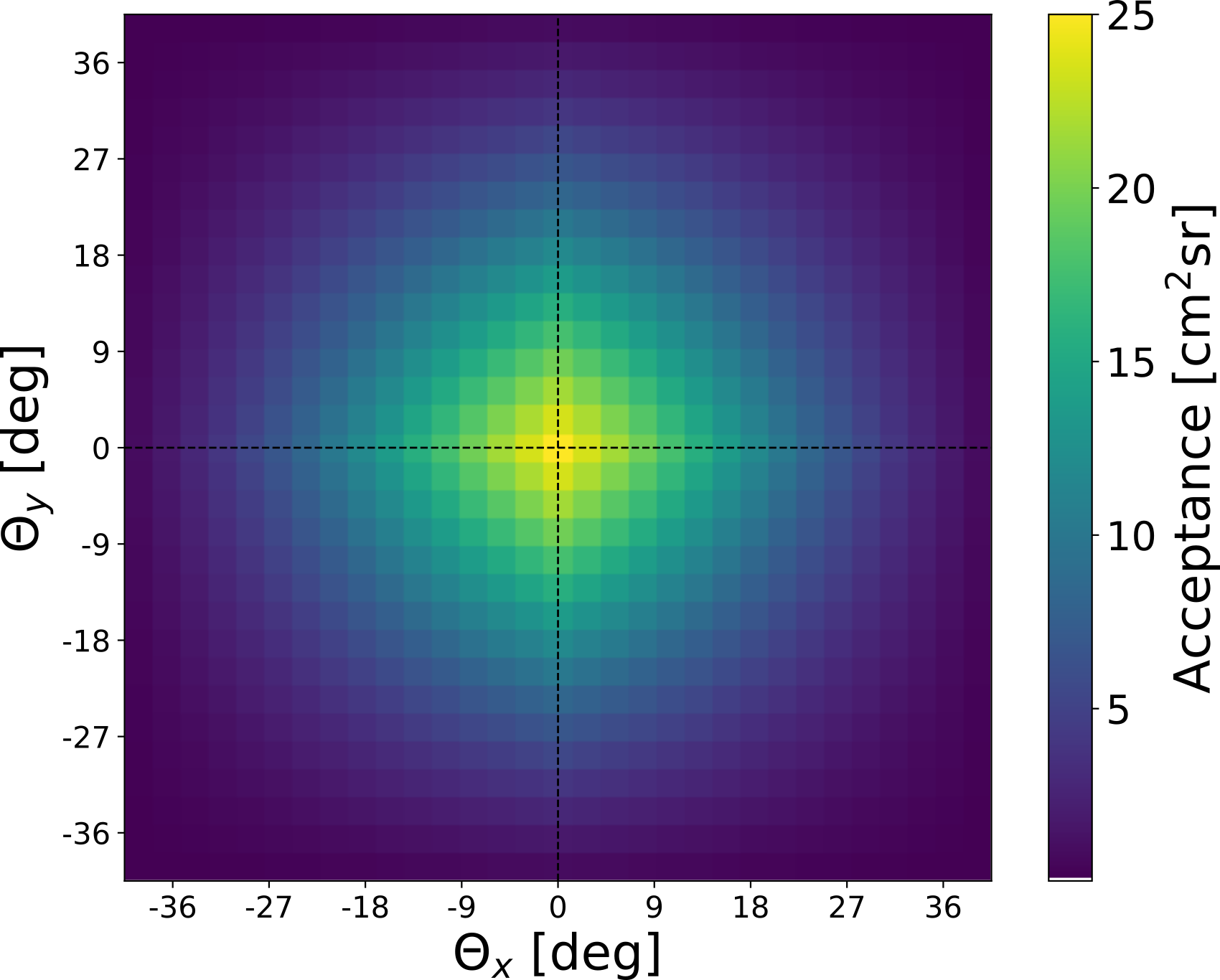} 
\includegraphics[width=.45\textwidth]{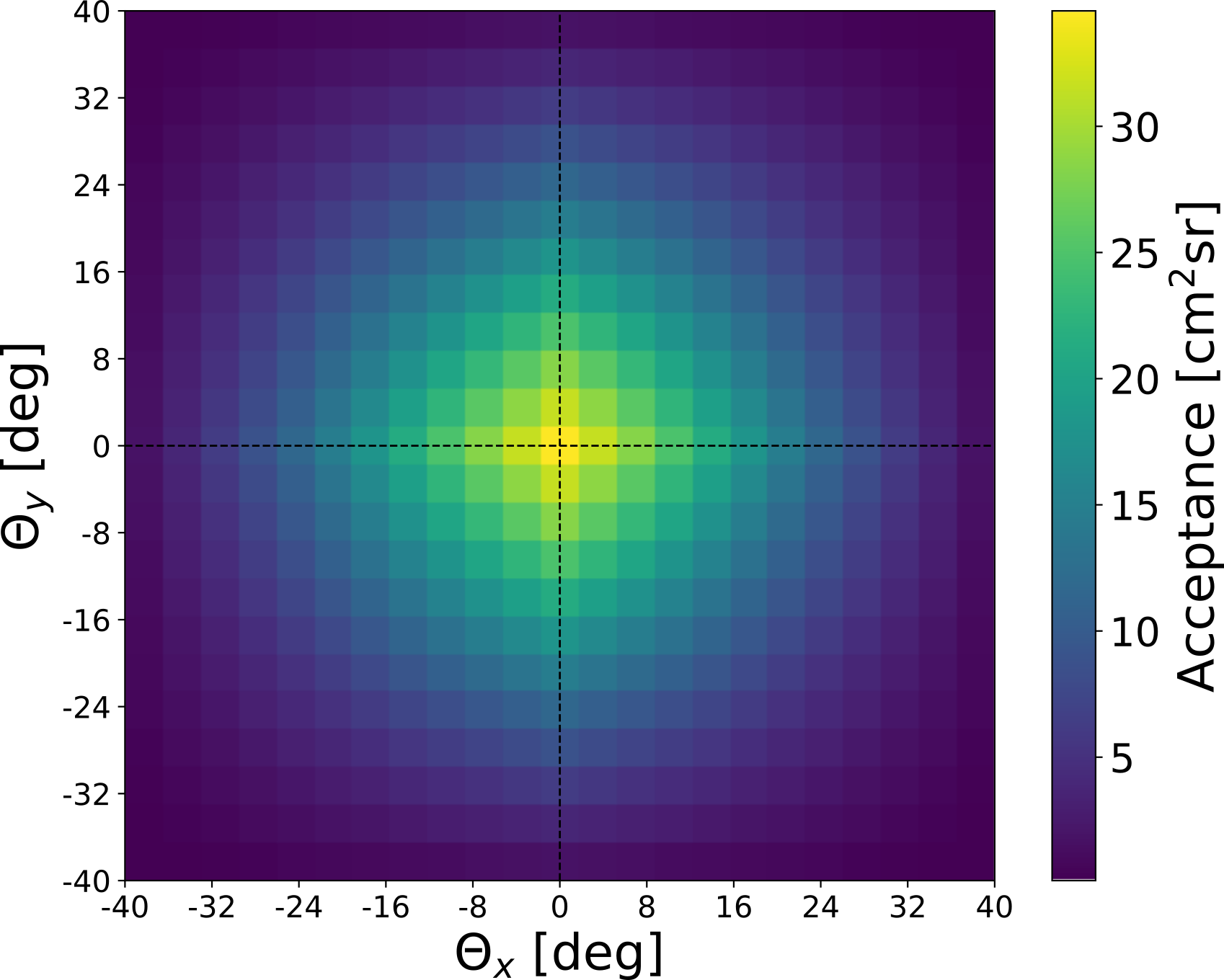}
\vfil}
\caption{Muon telescope acceptances. (Top-left) Gilbert et al. propose a telescope with a maximum acceptance in $r_{0,0}$ is 64 $cm^2 sr$. (Top-right) Carbone et al. detector having a full acceptance in $r_{0,0}$ gis 5.54 $cm^2 sr$. (Middle-left) Lo Presti et al. telescope with an angular aperture of $\pm$45$^{\circ}$, and a full acceptance in $r_{0,0}$ is 1.04 $cm^2 sr$. (Middle-right) Lesparre et al. design a telescope with a maximum acceptance in $r_{0,0}$ is 12.1 $cm^2 sr$, and an angular aperture of $\pm$34$^{\circ}$. (Bottom-left) After changing $D = 80 cm$, the maximum acceptance in $r_{0,0}$ is 25\,$cm^2 sr$, and an angular aperture of $\pm$45$^{\circ}$. (Bottom-right) Uchida et al. propose a telescope with a maximum acceptance in $r_{0,0}$ is 34.57 $cm^2 sr$, and an angular aperture is $\pm$ 40 $^{\circ}$.}
\label{Acceptance}
\end{figure*}

Figure \ref{Acceptance} reproduces results obtained by six muon telescopes. Gilbert et al. present a scintillator hodoscope composed of two matrices with $N_x = N_y = 32$ strips, a pixel side $d = 5$~cm, and a separation $D = 100$~cm \cite{Gibert2010}. The maximum acceptance in $r_{0,0}$ is 64\,cm$^2$~sr with an angular aperture of $\pm$57$^{\circ}$. Carbone et. al describes a telescope with $N_x = N_y = 16$ strips, a pixel side $d = 5$~cm, and a separation $D = 170$~cm \cite{Carnone2013}. The maximum acceptance in $r_{0,0}$ is 5.54\,cm$^2$sr. Lo Presti et al. describe a muon-tracking detector of two layers of $N_x = N_y = 99$ extruded plastic scintillator bars, a pixel side $d = 1$~cm, and a separation $D = 97$~cm \cite{LoPresti2020}. The angular aperture is about $\pm$45$^{\circ}$, the angular resolution has a maximum of 4.25$\times 10^{-4}$~sr and the maximum acceptance in $r_{0,0}$ is 1.04\,cm$^2$sr.   

\begin{figure*}
\centering
\vbox to130mm{\vfil 
\includegraphics[width=.45\textwidth]{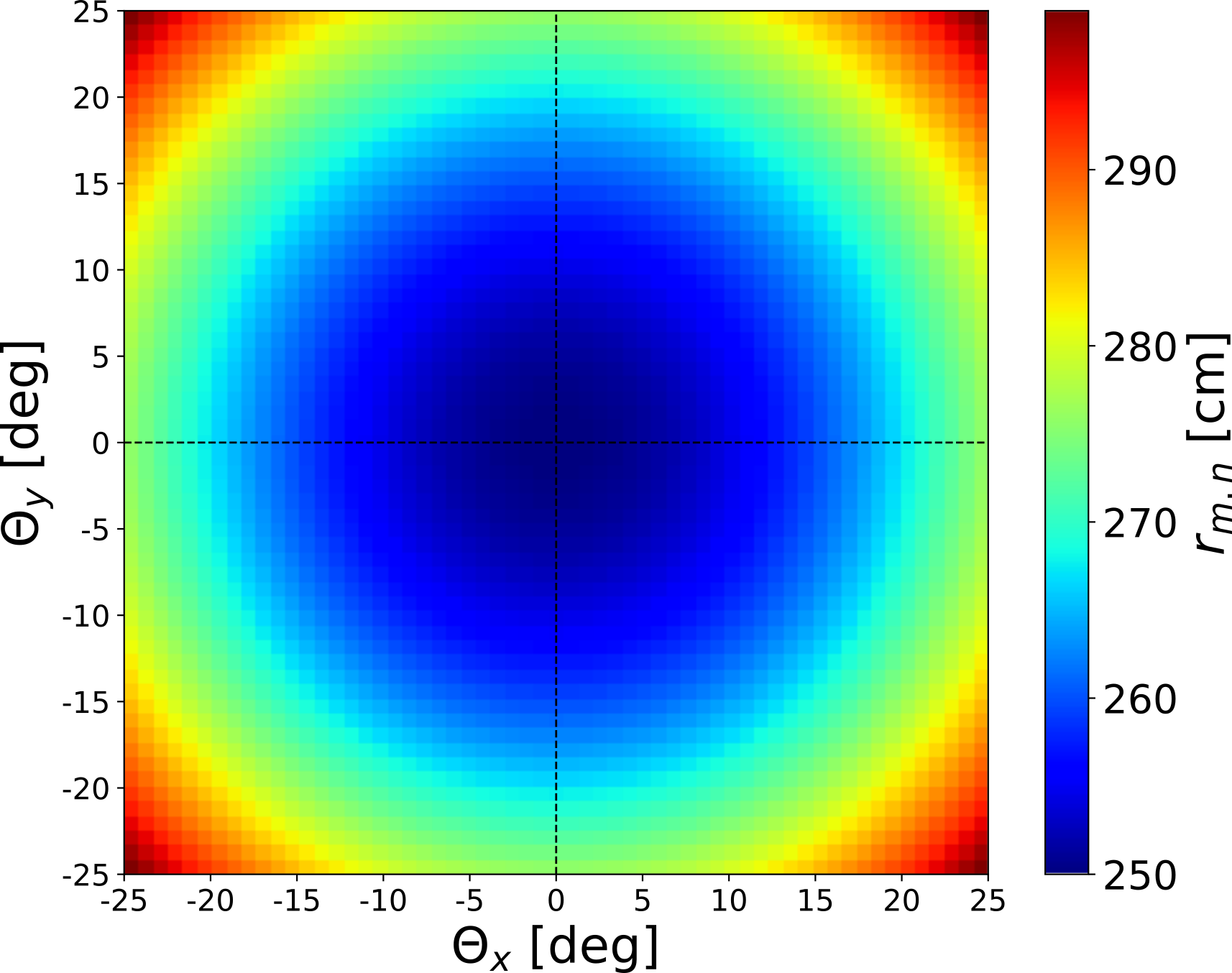}
\includegraphics[width=.45\textwidth]{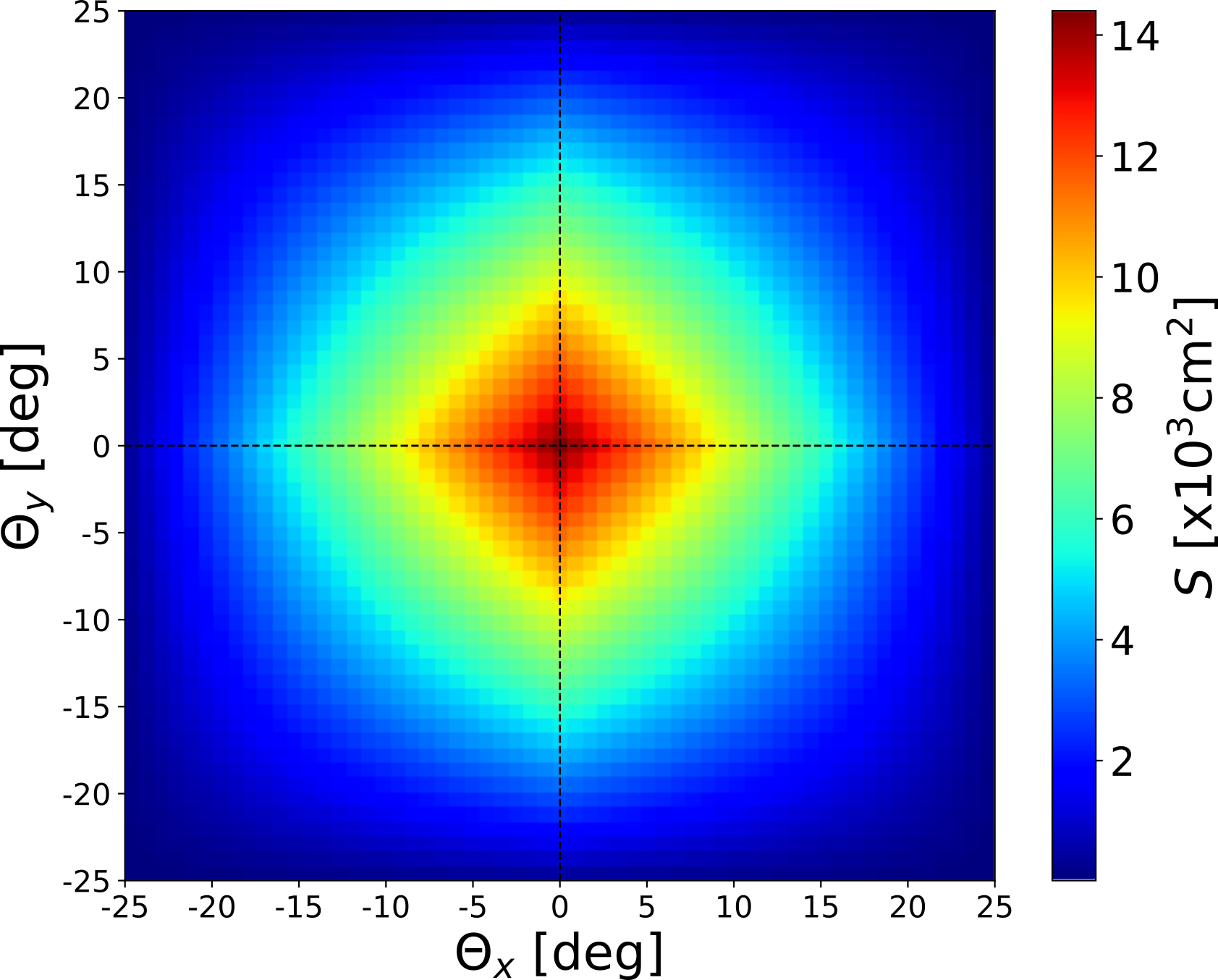} \\
\includegraphics[width=.45\textwidth]{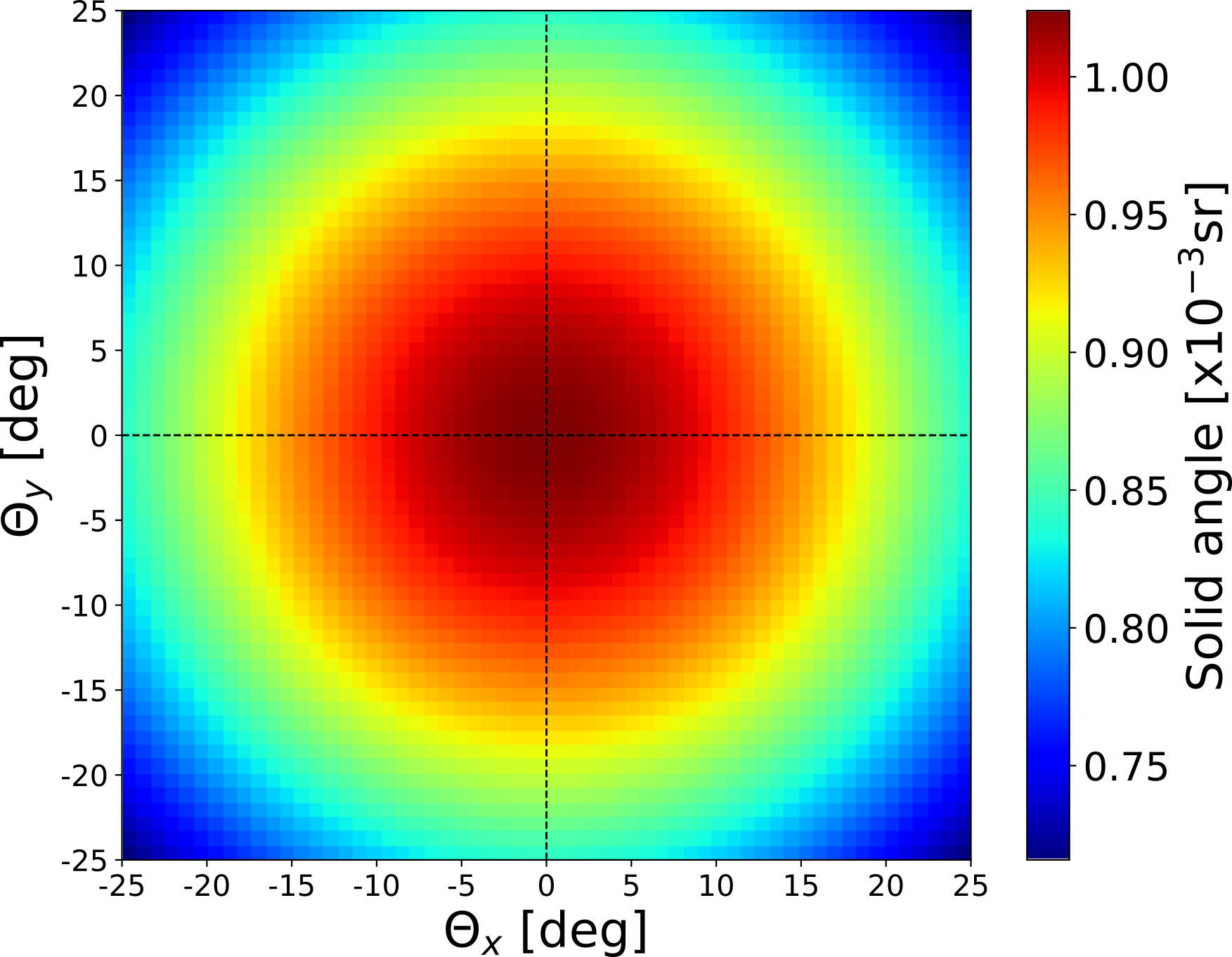}
\includegraphics[width=.45\textwidth]{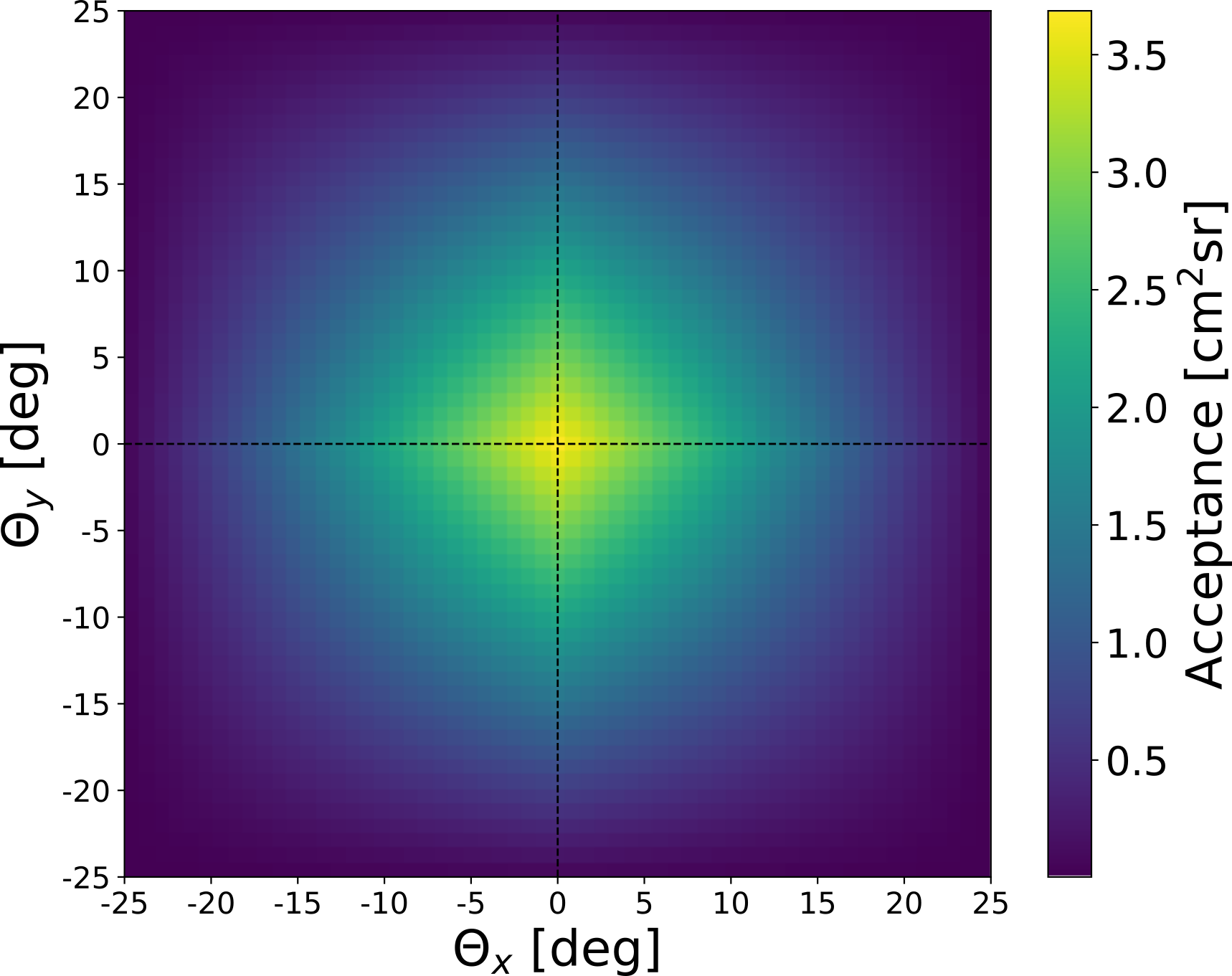}
\vfil}
\caption{Parameterization of the Muon Telescope for $N_x = N_y = 30$, a pixel side $d = 4$~cm, and a separation $D = 250$~cm. The MuTe maximum detection area is $\sim 14.4$~cm$^2$, the solid angle for perpendicular trajectories reaches $\sim$1.02$\times 10^{-3}$~sr, and the maximum acceptance is $\sim$3.6\,cm$^2$sr for $r_{0,0}$.}
\label{MuTe_params}
\end{figure*}

Lesparre et al.~\cite{Lesparre2012} designed a telescope with $N_x = N_y = 16$ strips, a pixel side $d = 5$~cm, and a separation $D = 115$~cm. The maximum acceptance in $r_{0,0}$ is 12.1\,cm$^2$sr, an angular aperture of $\pm$34$^{\circ}$ and an solid angle maximum of 7.56$\times 10^{-3}$~sr. The same telescope was set to $D = 80$~cm~\cite{Lesparre2010}. The maximum acceptance in $r_{0,0}$ is 25\,cm$^2$sr, an angular aperture of $\pm$45$^{\circ}$ and an solid angle maximum of 15.6$\times 10^{-3}$~sr. Uchida et al. describes a telescope of $N_x = N_y = 12$ strips, a pixel side $d = 7$~cm, and a separation $D = 100$~cm \cite{Uchida2009,Lesparre2010}. The angular aperture is about $\pm$40$^{\circ}$, the angular resolution has a maximum equal to 19.6$\times 10^{-3}$~sr and the maximum acceptance in $r_{0,0}$ is 34.57\,cm$^2$sr. 

This module outputs the telescope's parameters such as the detection area, the angular resolution, the solid angle, and the acceptance. Fig. \ref{MuTe_params} shows the Muon Telescope, MuTe, parameters for $N_x = N_y = 30$ scintillator strips, a pixel side $d = 4$~cm, and a separation $D = 250$~cm \cite{Rodriguez2020}. The MuTe maximum detection area is $\sim$14.4\,cm$^2$, the solid angle for perpendicular trajectories reaches $\sim$1.02$\times 10^{-3}$~sr, and the maximum acceptance is ($\sim$3.6\,cm$^2$sr) for $r_{0,0}$ \cite{Rodriguez2020}.

\section{Muon tomography}
\label{chap::tomography}
Muon tomography reconstructs three-dimensional density distributions using several two-dimensional muograms \cite{Nagahara2018}, estimating 3D density using reconstruction algorithms. A two-dimensional muogram integrates the object density along the muon path but does not distinguish density anomalies along the direction. There exist three families of reconstruction algorithms: analytical (e.g. Filtered Back Projection, FBP), statistical (e.g. Max Likelihood), and algebraic methods (e.g. Algebraic Reconstruction Techniques, ART). Density analysis can be done using techniques such as the inversion of gravimetry data but require a priori geological information or a combination with other geophysical data. Long exposure times, a low muon flux and given topography result in a low number of unequally spaced projections of the object. Such limitations define the appropriate tomography reconstruction approach. 

FBP is computationally efficient but requires many equally spaced projections, which are impossible in muon tomography. The low number of projections and long acquisition times limit three-dimensional density reconstruction of geological targets using muon tomography \cite{Barnoud2021}. Algebraic methods give good reconstruction results with a lower number of projections compared with FBP.

MUYSC handles the muon tomography reconstruction based on the Algebraic Reconstruction Technique. To implement muon tomography reconstruction in MUYSC, we evaluated the performance of various reconstruction algorithms found in the library the TomoPy library \cite{Gursoy2014}. The test consisted in performing a tomographic reconstruction on the Shepp-Logan object, with 180 observation points located uniformly between $[0,\pi]$ radians. The results and execution times are shown in \ref{app::perf}. It is clear from Fig. \ref{fig::Shepps_alg} that the best results are given by ART and Gridrec algorithms.

\begin{figure}
\begin{center}
\includegraphics[width=0.45\textwidth]{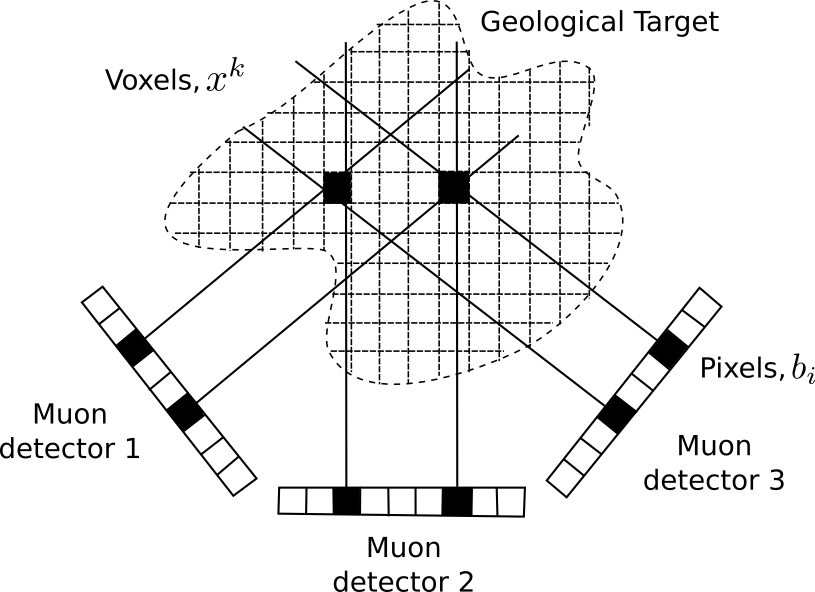}
\caption{Muon tomography based on an Algebraic Reconstruction Model. ART reconstructs a voxel ($x^k$) of the three dimensional structure from the projection data pixels ($b_i$) recorded by three different muon detectors. }
\label{fig::art}
\end{center}
\end{figure}

\subsection{Algebraic Reconstruction Technique}
An ART does not require an initial model as ordinary inversion analysis methods do~\cite{Nagahara2018}. This method uses a system of linear equations $Ax=b$ where $x$ is the volume vector to reconstruct, given the projection data $b$, with
\begin{equation}
    x^{k+1}=x^{k}+\lambda _{k}{\frac {b_{i}-\langle a_{i},x^{k}\rangle }{\|a_{i}\|^{2}}}a_{i}^{T},
\end{equation}
where $A$ is a sparse $m \times n$ matrix whose values represent the relative contribution of each output pixel to different points in the projection data ($m$ being the number of individual values in the projection data, and $n$ being the number of output voxels). As it is shown in figure \ref{fig::art}, each projection $b_i$ comprises several discrete values arranged along the transverse axis \cite{Herman2009}.

\subsection{Three-dimensional reconstruction with MUYSC}
We tested the MUYSC tomography module using simulated data from the MUYSC muon radiography module. It allows us to generate projections of the target from several points, such as to evaluate the 3D reconstruction performance. MUYSC simulated a muon tomography of the Cerro Machín Volcano with 42 observation points at 1059\,m $\pm$40\,m. from the interesting point with a mean altitude of 2495\,m a.s.l. $\pm$10\,m. All points were located in topography-suitable places around the volcano dome. 

Each data projection ($100\times100$ pixels) contains information on the muon path in a zenith range of 28$^{\circ}$ and azimuth range of 100$^{\circ}$. The tomography module of MUYSC gives a three-dimensional density distribution of the Cerro Machin volcano ($100\times100\times100$ voxels). 

The systematic error $\varepsilon$ assesses the reconstruction performance as follows \cite{Nagahara2018}, 
\begin{equation}
    \varepsilon_{i,j} =  R_{i,j}  - T_{i,j} ,
\end{equation}
$R$ is the sum of perpendicular slices across a given volume axis and $T$ is the original projection data on the axis. 

\begin{figure}[h!]
\begin{center}
\includegraphics[width=0.43\textwidth]{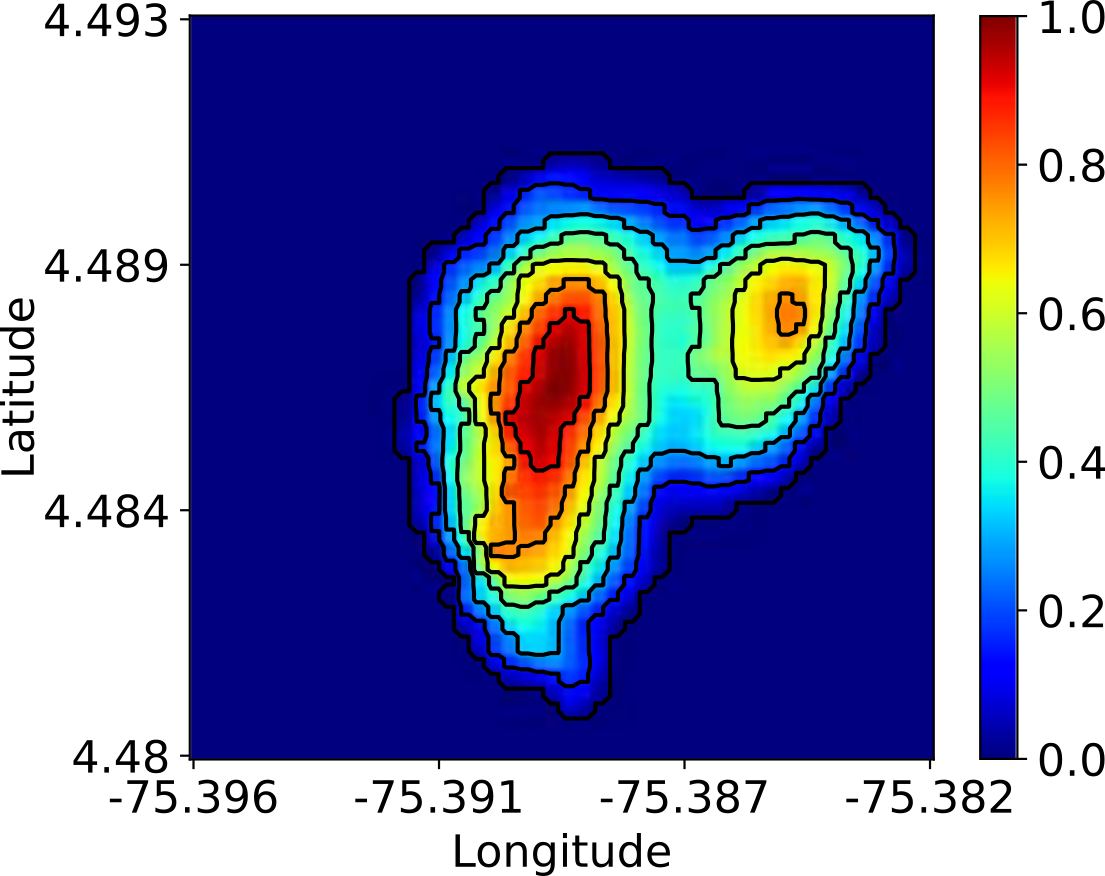} 
\includegraphics[width=0.43\textwidth]{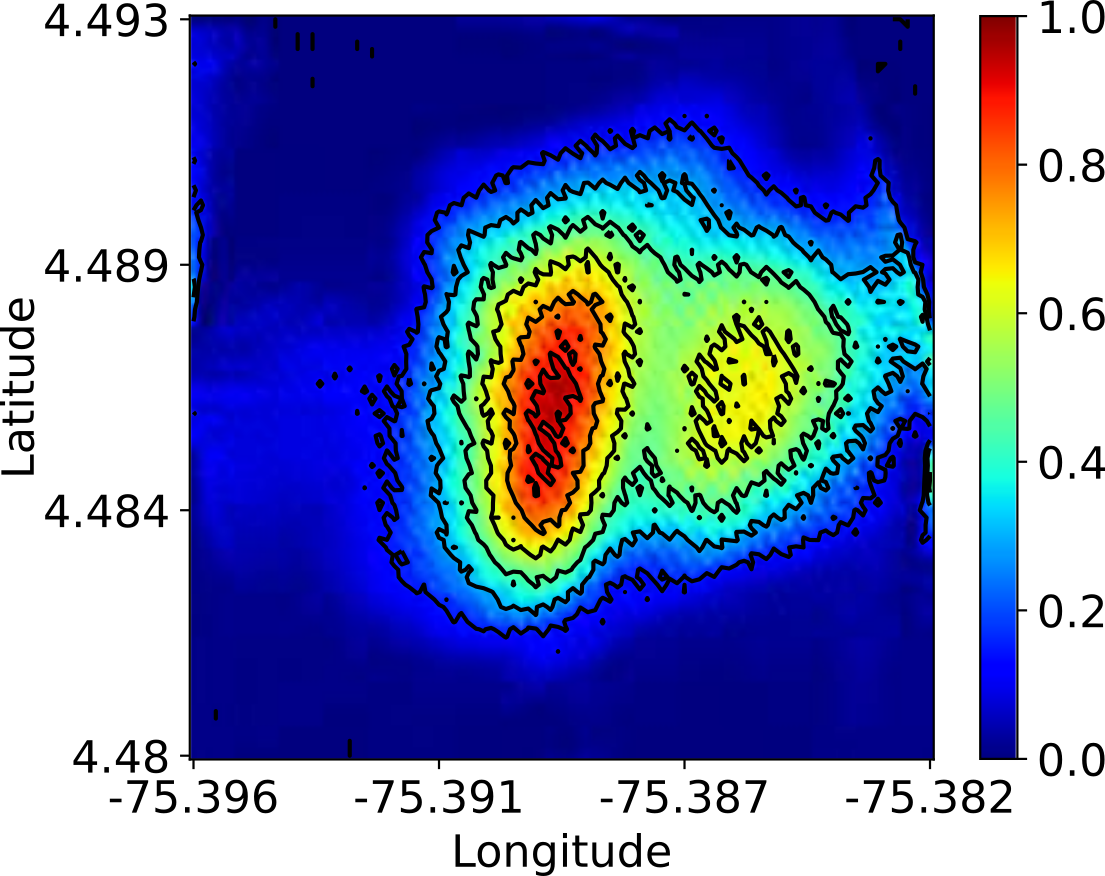}
\includegraphics[width=0.43\textwidth]{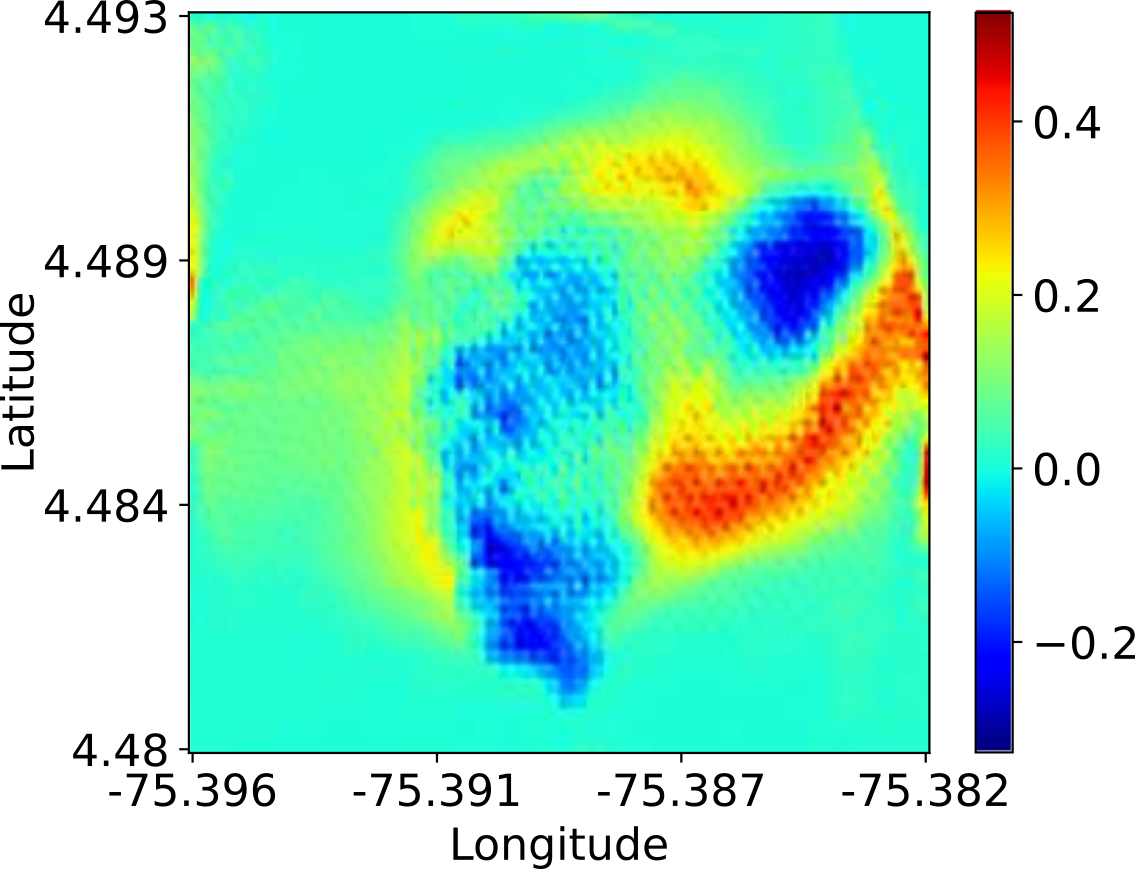}
\caption{Top view of the original structure (top), the reconstructed volume (middle), and the systematic error map (bottom). The topography cut ranges 4.479$^{\circ}$ to 4.494$^{\circ}$ Latitude and -75.396$^{\circ}$ to -75.381$^{\circ}$ Longitude. The secondary volcano dome reconstructed by MUYSC got a higher systematic error variance (0.825) in comparison with the principal reconstructed dome.}
\label{fig::Recons_Top}
\end{center}
\end{figure}

For the top view, It is necessary to estimate the reconstruction region. This is done by converting the angular view of the telescope to latitude, Longitude and height. Fig. \ref{fig::Recons_Top} shows the original and reconstructed volume with normalized units. The error map shows a minimum value ($\varepsilon_{\min} = -0.327$) on the top of the secondary dome and a maximum value ($\varepsilon_{\max} = 0.525$) around it, while an error on the central dome reaches ($\varepsilon_{\min} = -0.309$). Note that the error variation on the main dome is lower than in the secondary.

We calculated the error average $\mu_{\varepsilon}$ and the standard deviation $\sigma_{\varepsilon}$ by taking the absolute value of $\varepsilon$ in the volcano region. We got an error average of $\mu_{|\varepsilon|}=0.105$, with a standard deviation of $\sigma_{|\varepsilon|}=0.099$.

\begin{figure}[h!]
\begin{center}
\includegraphics[width=0.43\textwidth]{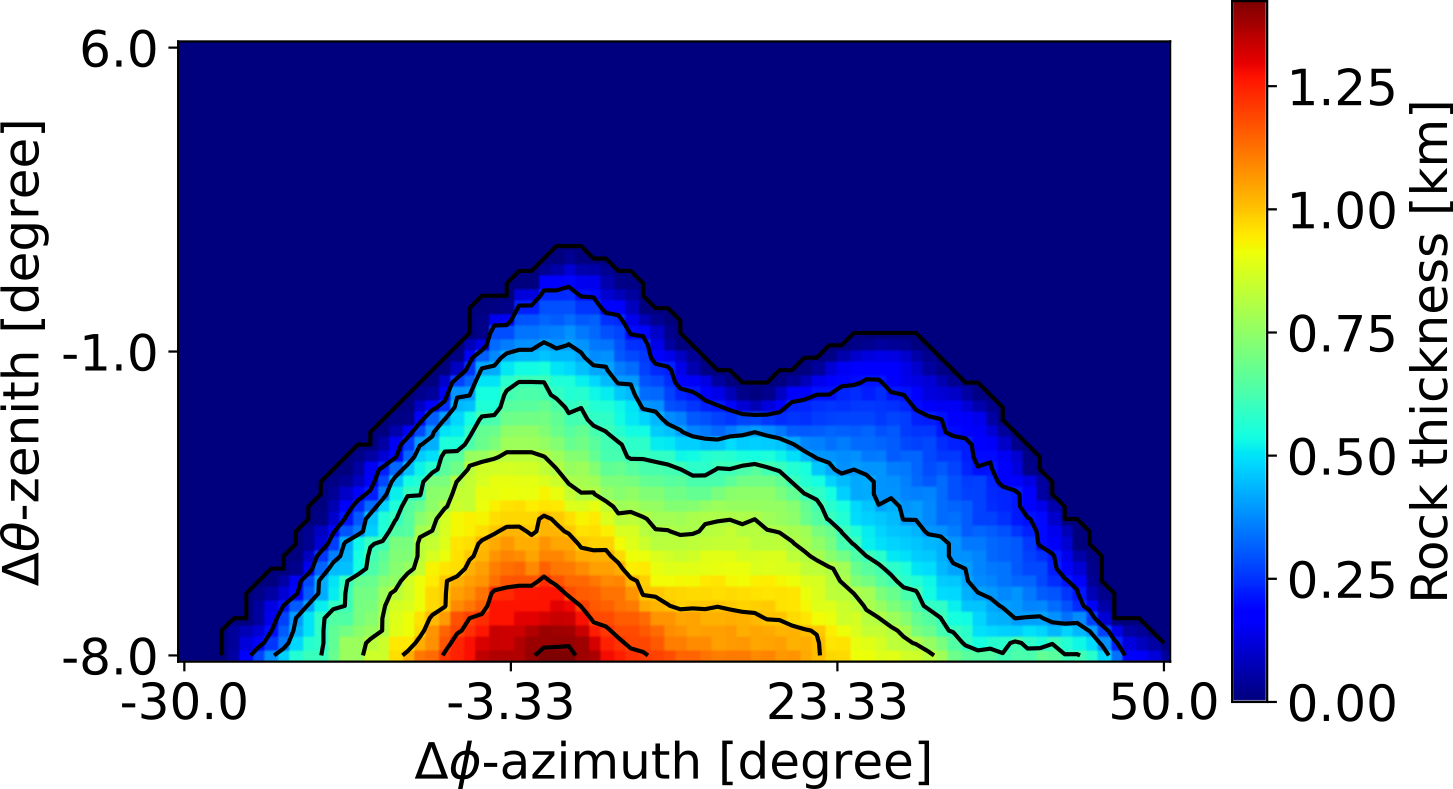} 
\includegraphics[width=0.43\textwidth]{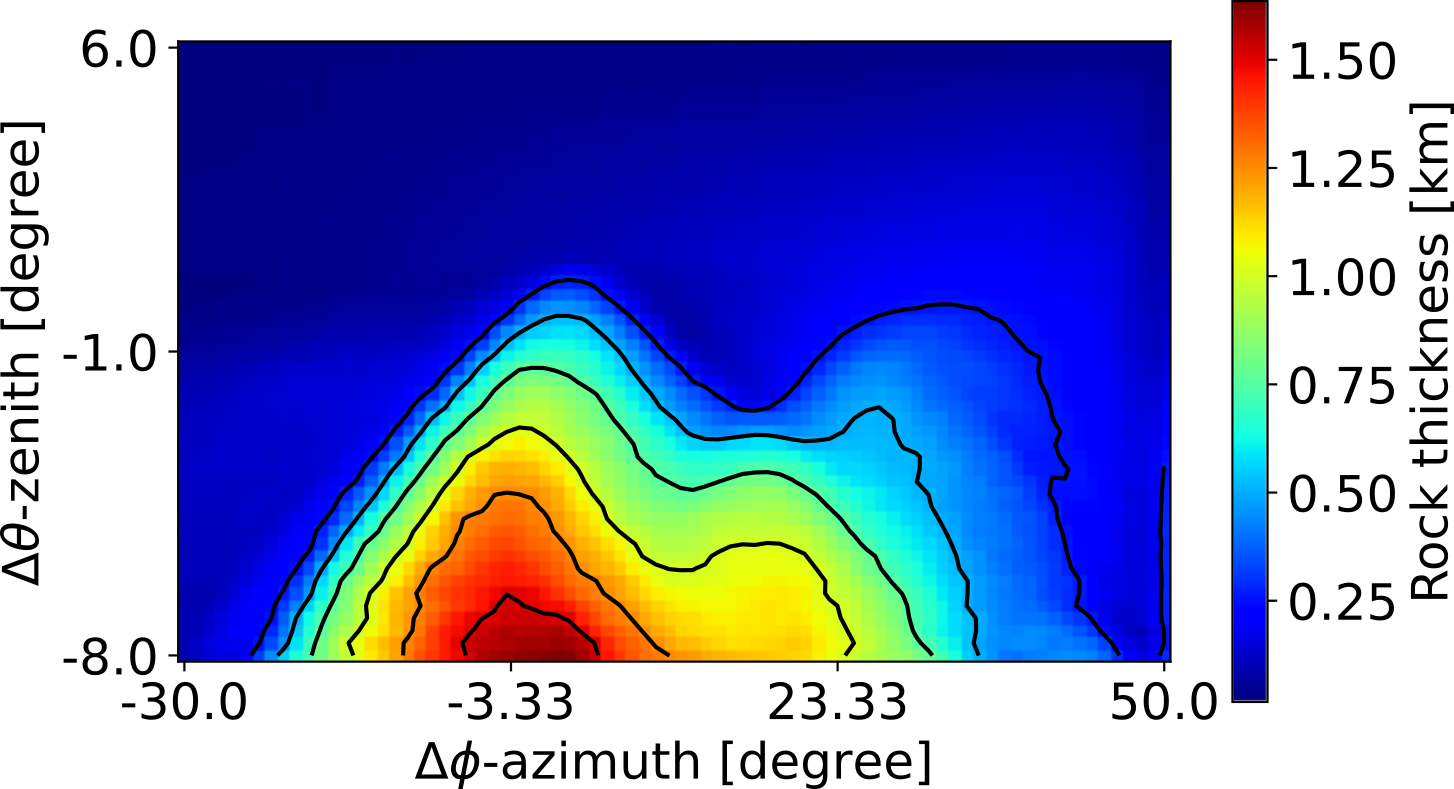} \\
\includegraphics[width=0.43\textwidth]{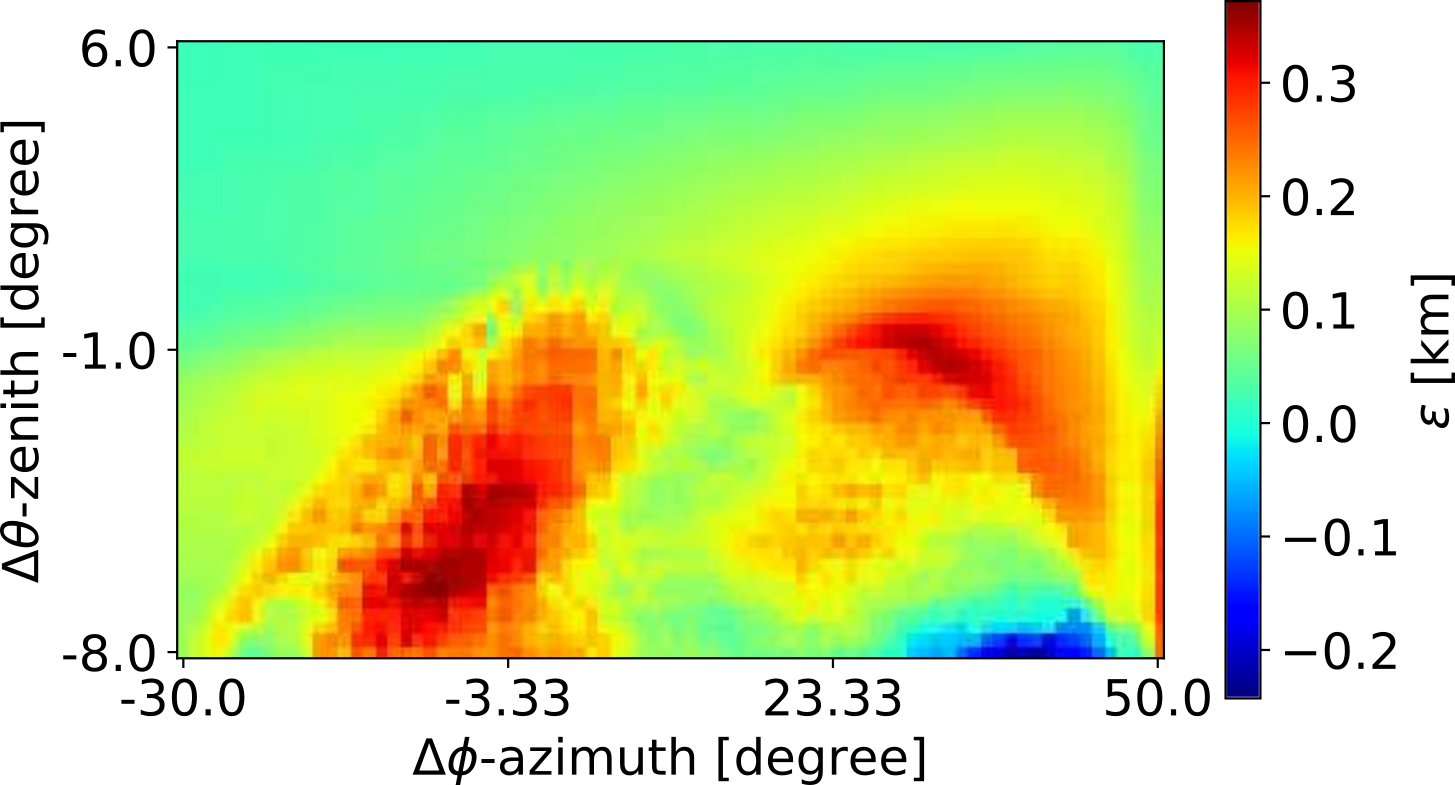}
\caption{Frontal view of the original structure (top), the reconstructed volume (middle), and systematic error map (bottom). The topography cut ranges -30$^{\circ}$ to 50$^{\circ}$ Azimuth and -8$^{\circ}$ to -6$^{\circ}$ from the principal dome peak. The systematic error reaches 0.37\,km in the middle of the principal volcano dome and at the top of the secondary dome.}
\label{fig::Recons_Front}
\end{center}
\end{figure}

For the anterior view, we used a projection from the front of the structure (0.017\,rad). Fig. \ref{fig::Recons_Front} shows the original and reconstructed volume in units of rock thickness. MUYSC fully reconstructs the volcano shape. The systematic error increases in the middle of the principal dome and at the top of the secondary one. The error reaches ($\varepsilon_{max} = 0.37$ km), with $\mu_{|\varepsilon|}=0.158$ km and $\sigma_{|\varepsilon|}=0.0768$ km.

The three-dimensional reconstruction depends strongly of the number of projections. The lower the number of projections, the lower the reconstruction performance.

\section{Conclusions}
\label{chap::conclusions}
This work describes MUYSC, a software tool to simulate muon radiography and muon tomography of geological objects. MUYSC simulates muography measurements as follows:

\begin{itemize}
    \item the integrated muon flux passing across the geological object
    \item the minimum exposure time to get given muon per pixel threshold
    \item the geological object transmission matrix
    \item the geological object opacity
\end{itemize}
 
We reproduced the results of different muography experiments using the MUYSC framework. MUYSC got rock thickness and integrated muon flux traversing the following volcanoes: Cerro Machín volcano \cite{VesgaRamirez2020}, Mount Etna \cite{Carnone2013}, the Puy de Dôme \cite{Menedeu2016}, and the Mount Vesuvius \cite{Enricco2022}, taking into account the observational conditions (geographical coordinates and detector elevation) reported by the authors. The high computational performance of MUYSC allows it to solve muon radiography (100 $\times$ 100 pixels) of a volcanic structure in no more than 2 minutes.

MUYSC also calculates the telescope parameters as follows:

\begin{itemize}
    \item the angular resolution
    \item the acceptance
    \item the distance matrix of particles crossing along the detector
\end{itemize}

For different detector configurations (inter-panel distance or pixel size), we computed the muon detector acceptance and solid angle of eight muon detectors changing the number of pixels and inter-panel distances \cite{Gibert2010,Carnone2013,LoPresti2020,Lesparre2012,Lesparre2010,Uchida2009}. 

The MUYSC tomography module evaluates the viability of carrying out a three-dimensional density reconstruction of a geological object from a given number of muon radiography projections. We presented the reconstruction of the Cerro Machin volcano from 42 muon radiography projections and the systematic error evaluation.

MUYSC runs on computing facilities from personal notebooks to high-performance clusters. The MUYSC code can be easily obtained from a GitHub repository through the Muon Telescope\footnote{\url{https://halley.uis.edu.co/fuego/en/muysc-2/}} web page. The documentation (installation and user manual) is also available. 

*MUYSC has the advantage of being a specialized/optimized software tool for muography, reducing computational time and boosting the expected results. MUYSC does not pretend to replace software tools such as CORSIKA or GEANT4.
\begin{acknowledgments}
We gratefully acknowledge the support of the Vicerrector\'ia de Investigaci\'on y Extensi\'on from Universidad Industrial de Santander under project VIE2814.
\end{acknowledgments}

\bibliographystyle{gji}
\bibliography{Biblio.bib}

\appendix

\section{Performance of various reconstruction algorithms}
\label{app::perf}
\begin{figure*}
\begin{center}
\vbox to80mm{\vfil
\includegraphics[width=0.85\textwidth]{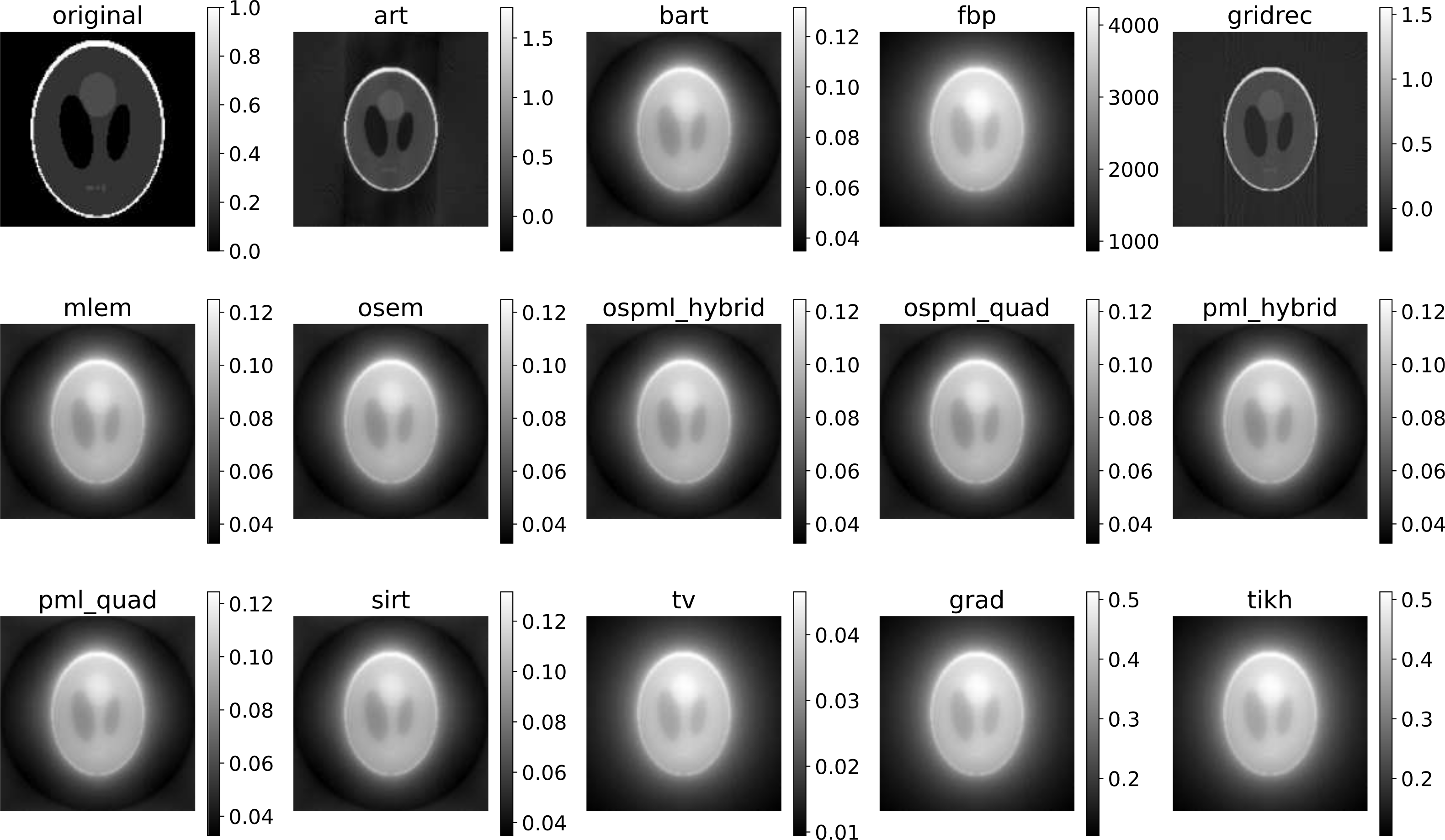}
\vfil}
\caption{Reconstruction of Shepp-Logan object with different algorithms. }
\label{fig::Shepps_alg}
\end{center}
\end{figure*}

\begin{table*}
\begin{minipage}{106mm}
\caption{Execution time of TomoPy reconstruction algorithms. }
\label{tab::Exec_time}
\begin{tabular}{|p{0.6\textwidth}|p{0.1\textwidth}|}
\hline
\textbf{Algorithm}                                                                                                 & \textbf{Execution time (s)} \\ \hline
Algebraic reconstruction technique (ART)                                                                           & 2.56812                     \\ \hline
Block algebraic reconstruction technique (BART)                                                                    & 12.1972                     \\ \hline
Filtered back projection (FBP)                                                                                     & 8.64134                     \\ \hline
Fourier grid reconstruction algorithm (Gridrec)                                                                    & 0.13586                     \\ \hline
Maximum-likelihood expectation maximization algorithm (ML-EM)                                                      & 11.41385                    \\ \hline
Ordered-subset expectation maximization algorithm (OSEM)                                                           & 10.04587                    \\ \hline
Ordered-subset penalized maximum likelihood algorithm with weighted linear and quadratic penalties (OSPML\_hybrid) & 10.73985                    \\ \hline
Ordered-subset penalized maximum likelihood algorithm with quadratic penalties (OSPML\_quad)                        & 10.90423                    \\ \hline
Penalized maximum likelihood algorithm with weighted linear and quadratic penalties (PML\_hybrid)                  & 10.64516                    \\ \hline
Penalized maximum likelihood algorithm with quadratic penalty (PML\_quad)                                          & 13.03399                    \\ \hline
Simultaneous algebraic reconstruction technique (SIRT)                                                             & 11.28778                    \\ \hline
Total Variation reconstruction technique (TV)                                                                      & 10.93927                    \\ \hline
Gradient descent method (Grad)                                                                                     & 10.81452                    \\ \hline
Tikhonov regularization with identity Tikhonov matrix (Tikh)                                                       & 12.60597                    \\ \hline
\end{tabular}
\end{minipage}
\end{table*}

\label{lastpage}

\end{document}